\documentclass[useAMS,usenatbib]{mn2e}
\usepackage{graphicx}
\usepackage{caption}
\usepackage{epsfig}

\usepackage{blindtext}

\newcommand{\fe}[0]{[{\rm Fe/H}]}

\newcommand{\kpc}[0]{{\rm \, kpc}}

\newcommand{\vlos}[0]{v_{\rm los}}

\newcommand{\vmu}[0]{{\bf v}_\mu}

\newcommand{\vdep}[0]{\pi^{-1}(\vlos)} 

\def\mag{{\rm \, mag}}

\def\kms{{{\rm \, km}s^{-1}}}

\title[Rotational signature of the Milky Way stellar halo]{Rotational signature of the Milky Way stellar halo}

\author[Fermani \& Sch\"onrich]{Francesco Fermani$^{1}$\thanks{E-mail:f.fermani1@physics.ox.ac.uk}, Ralph Sch\"onrich$^{2}$\\
$^{1}$University of Oxford, Rudolf Peierls Centre for Theoretical Physics, Oxford OX1 3NP\\
$^{2}$Hubble Fellow, Department of Astronomy, Ohio State University, Columbus, USA}

\begin{document}

\date{Accepted 2013 April 9.  Received 2013 April 8; in original form 2012 July 5}

\pagerange{\pageref{firstpage}--\pageref{lastpage}} \pubyear{2013}

\maketitle

\label{firstpage}

\begin{abstract}
 We measure the rotation of the Milky Way stellar halo on two samples of Blue Horizontal Branch (BHB) field halo stars from the Sloan Digital Sky Survey (SDSS) with four different methods. The two samples comprise 1582 and 2563 stars respectively and reach out to $\sim 50 \kpc$ in galactocentric distance. Two of the methods to measure rotation rely exclusively on line-of-sight velocities, namely the popular double power-law model and a direct estimate of the de-projected l.o.s. velocity. The other two techniques use the full 3D motions: the radial velocity based rotation estimator of \cite{SBA} and a simple 3D azimuthal velocity mean. In this context we a) critique the popular model and b) assess the reliability of the estimators. All four methods agree on a weakly prograde or non-rotating halo. Further, we observe no duality in the rotation of sub-samples with different metallicities or at different radii. We trace the rotation gradient across metallicity measured by \cite{de11} on a similar sample of BHB stars back to the inclusion of regions in the apparent magnitude-surface gravity plane known to be contaminated. In the spectroscopically selected sample of \cite{x11}, we flag $\sim 500$ hot metal-poor stars for their peculiar kinematics w.r.t. to both their cooler metal-poor counter-parts and to the metal-rich stars in the same sample. They show a seemingly retrograde behaviour in line-of-sight velocities, which is not confirmed by the 3D estimators. Their anomalous vertical motion hints at either a pipeline problem or a stream-like component rather than a smooth retrograde population.
\end{abstract}


\section{Introduction}
The stellar halo represents an observable tracer of the putative dark matter halo which is thought to govern our cosmological evolution \citep[e.g.][]{WR}. The present observational evidence favours a hierarchical formation scenario, where the stellar halo was formed through accretion of protogalactic fragments \citep[][]{se78,Bull05}, especially after the detection of abundant structures with SDSS \citep[][]{be08,bel06}. Simulations of stellar halos support the picture of a halo mostly made of accreted structures \citep[e.g.][]{co10}, but \cite{free11} remarks that a formation process via dissipation cannot be completely ruled out \citep[][]{egg62,sam03}.

A transition from a flattened inner to a more spherical outer halo has sometimes been traced back to halo formation in different stages and via different physical processes \citep[][]{hart87,pres91}. Since it has been argued that dynamical friction is more efficient for prograde infall \citep[][]{qui86,byr86} there is a popular expectation for the outer halo to have a different rotational signature from the inner halo \citep[][]{mur10}, and in particular to be counter-rotating \citep[][]{nor89}.

The kinematics of the stellar halo could serve as a probe of formation history, but it is still highly controversial: on the one hand there are claims for a globally counter-rotating halo \citep[e.g.][]{maj92}. More recently \cite{car07,car10}  using solar neighbourhood main sequence stars claimed a inner pro-rotating halo opposed to an outer counter-rotating halo, which \cite{kin07,kin12} confirmed with a small sample of RR Lyrae stars and blue horizontal branch
 stars localized near the Galactic anticentre direction. With a larger sample of blue horizontal branch field halo stars \cite{de11} found instead a metal-rich pro-rotating halo opposed to a metal-poor counter-rotating halo with such a dichotomy being coherent at all radii. On the other hand \cite[][]{chi00} and \cite{sir04} argue for a non-rotating halo, but do not distinguish for different metallicity.

The uncertainty in this debate arises from  several causes: the challenge of a clean selection of halo stars, the question if such samples give a fair representation of the halo (e.g. still debated for globular clusters or the metallicity biases in red or blue giant stars), as well as uncertainties in determining uniquely their kinematics. Stars in the solar neighbourhood have well-determined kinematics \citep[e.g. GCS, RAVE: ][]{nor04,ste03}, however, they contain hardly any halo stars prohibiting an uncontaminated and unbiased halo sample \citep[][]{kin95}. To explore the halo in situ\footnote{By in situ we refer to stars that are actually observed there, rather than stars which are extrapolated to be there via orbit-integration.}, intrinsically brighter objects are needed \citep[][]{san70}, but for those mainly line-of-sight velocities are available and many distance schemes associated with these stars present metallicity dependent systematics which can produce a false gradient in kinematics across metallicity bins \citep[][]{P1}. Among the most popular tracers are BHB stars \citep[][]{sir04}, RR Lyrae stars and globular clusters \citep[e.g.][]{ode01}.

Distances for these stars are obtained spectro-photometrically: this implies that the uncertainty on the distance correlates with how noisy is the spectrum of a star. To assign an absolute magnitude to the star one needs to identify its luminosity type \citep[][]{bee00} and therefore to have an accurate assessment of its surface gravity $\log g$. This danger of misclassification is especially an issue for the noisy spectra of remote and hence faint objects deeper in the halo \citep[e.g.][]{yan00, sir04}. 
For example, parameters in automated pipelines can be drawn into unphysical values or into points that are a local minimum by averaging over the most likely points, in case the pdf is not unimodal \citep[see \S 4.4.3,][]{lee08a}.

It is therefore imperative to have a complete understanding of how such uncertainties propagate in our analysis and thus to bring possible biases under control: remarkable efforts have been made in this direction, but a complete set of rigorous methodologies has not been compiled yet. For example \cite{rya92} demonstrated that errors in proper motion due to overestimates in distance can easily bias the kinematic measurements towards a counter-rotating trend. \cite{SBA}, hereafter SBA12, lift the need to assume the shape for the velocity ellipsoid when estimating the distance of solar-neighbourhood stars by combining proper motion and line-of-sight velocity ($\vlos$): they achieve this by exploiting correlations between the measured $U$, $V$, $W$ components of velocity introduced by systematic distance errors and iteratively correcting the distance. \cite{SAC}  showed that the claim of a retrograde outer-halo in \cite{car07,car10} was a consequence of inappropriate treatment of distance errors.

The aim of this paper is to achieve similar clarity in the context of BHB halo stars where only $\vlos$ and tentative proper-motion information are available. Optimal control of systematics is essential to obtain a meaningful result in this sample. We attempt to meet this goal by contrasting two independent sample selections and four different estimators for halo rotation.

In Section 2 we introduce this set of rotation estimators, and assess the reliability of each method depending on whether they use $\vlos$ information only or the full 3D motion. Section 3 describes the two selections of BHB stars that we use as tracers of the stellar halo, and Section 4 presents the application of the rotation estimators on these samples. Section 5 compares our results with previous works, while Section 6 critiques the sample of X11 in detail. We sum up and look to the future in the Conclusions.

\section{Methodology}

We critique the model used by D11. Assuming a radial power law potential and a distribution function D11 estimates the rotation and anisotropy of the stellar halo from a sample of BHB field halo stars for which only $\vlos$ is available by marginalizing over proper motions. As an alternative, we assess the possibility of directly estimating rotation using either $\vlos$ only but also full 3D motion. The latter estimators extract the average rotation and therefore are suitable even at large proper motion errors, as long as the systematic bias is small. We conclude the section by recalling general principles of consistency for the rotational signature of a smooth component: these will later be useful to test the results of our analysis.

\subsection{Likelihood analysis}
One way to extract morphological and kinematic information from a sample is to fit all the data to a sufficiently general theoretical model. Commonly studied parameters of the Milky Way stellar halo are the rotation and the deviation of the velocity ellipsoid from a spherical shape, usually with the radial velocity dispersion varied against the two other directions. A generally accepted method is to assume a density profile for the stellar component and an overall potential for the Galaxy. Then one builds up a phase-space distribution function (DF) in the steady-state approximation, which will depend on parameters identifiable with anisotropy and rotation. One either can marginalise over observables that are considered too noisy or, more properly, convolve the distribution function with an error function replicating the observation process. To estimate a value for the physically meaningful parameters, one can look at the likelihood of the data given the model, varying the free parameters. The results must be robust to the uncertainty in the data and give consistent results for different subsets, and the model needs to be flexible enough to fit various behaviours and trends.

D11 also used BHB stars drawn from SDSS (DR7) to study the kinematics of the Milky Way stellar halo: thus, we adopt their model first to ensure consistency of our results. 

\subsubsection{Methodology of Deason et al. 2011}\label{sec:model}

D11 assume a radial potential of the type $\Phi \sim r^{-\gamma}$, normalized so that the escape velocity at the Sun is the one estimated by \cite{smi07}\footnote{The Sun is taken to be at a distance of 8.5 kpc from the Galactic Center. The escape velocity reported by \cite{smi07} is $v_{\rm esc}=498-608$ km s$^{-1}$ with $90 \%$ confidence and an average of $544$ km s$^{-1}$. We adopt the average value.}. The density profile of BHB stars is assumed to be also of power-law type: $\rho \sim r^{-\alpha}$. 

To estimate the rotational signature and anisotropy of the sample, they adopt a double power-law distribution function depending on two parameters: $\beta$ \cite[anisotropy parameter as defined in][ \S 4.3.2]{BT08} and $\eta$ (rotation), such that $\eta=0$ corresponds to complete pro-rotation and $\eta=2$ to complete counter-rotation:

\begin{equation}\label{model_df}
F_{\eta,\beta}(E,L,L_z) =c_\beta\left(1+(1-\eta)\tanh\left(\frac{L_z}{\Delta}\right)\right) L^{-2\beta}E^{\zeta},
\end{equation}

\noindent where $E$ is the binding energy, $\Delta$ is a smoothing parameter, $c_\beta$ is a constant that enables us to make the real-space density independent of $\beta$ and

\begin{equation}
\zeta=\frac{\beta(\gamma-2)}{\gamma}+\frac{\alpha}{\gamma}-\frac{3}{2}.
\end{equation}

Assuming dominant errors on the proper motions, the DF is marginalized over transverse velocities. The free parameters $(\eta,\beta)$ are then estimated via Markov Chain Monte Carlo (MCMC) sampling on the parameter space with the probability density being the likelihood of the data\footnote{Heliocentric galactic polar coordinates are converted into galactocentric cylindrical coordinates reference frame, by assuming $(x,y,z)_\odot = (8.5,0,0)\kpc$, $v_{c \odot}=220$ km s$^{-1}$. We are aware that this specific parameter combination is at odds with the measured proper motion of Sgr ${A^*}$, but keep those values consistent with previous studies. We further assume a solar Local Standard of Rest velocity of $(U,V,W)_\odot$=(11.1,12.24,7.25) km s$^{-1}$, as updated by \cite{SBD}. } given the model, so the probability density in parameter space is the exponential of:

\begin{equation}\label{MLH}
L(\eta,\beta)=\sum_{i=0}^{N} \log F_{\eta,\beta}(l_i,b_i,d_i,v_{{\rm los},i}),
\end{equation}

\noindent where 

\begin{equation}
F_{\eta,\beta}(l,b,d,\vlos)=\int \int \textrm{d}v_l \textrm{d}v_b F_{\eta,\beta}(l,b,d,v_l,v_b,\vlos).
\end{equation}

See Appendix A for details on the convergence of the MCMC estimate and for a test of the algorithm implementation.

\subsubsection{Notes on the method}

The choice of an index $\alpha \ge 3$ for the density profile makes it impossible to compute a normalization constant for the DF in (\ref{model_df}) as it contains an infinite mass at the centre: $\int \textrm{d}^3x \textrm{ d}^3 v \textrm{ }F_{\eta,\beta}(E,L,L_z)= \infty$. Therefore, in order to be able to compare different models we must fix the density in the sampled region of real space. Thus

\begin{equation}
\rho(r)=\int   \textrm{d}^3 v \textrm{ } F_{\eta,\beta}(E,L,L_z)
\end{equation}
is independent of $\beta$. 

A convenient consequence of fixing the density is that we can neglect selection effects which would otherwise apply \citep[][]{mcm11}. All the same, we are left with the uncertainty on the  real density profile of the tracers. If the steepness of the profile is overestimated, for example, the model considers distant stars less probable than they really are. This results in these objects having more weight than their closer companions. An unfortunate aggravation of this bias is that fainter/more remote stars are known to be noisier and are intrinsically a bad tracer of the rotational signature.

When marginalizing over proper motions with $\beta >0$, we compute an integral that involves a singularity in the DF (\ref{model_df}) at $L=0$ and the treatment of this singularity is crucial to retrieve $\beta$ correctly. 

Fortunately, the singularity is integrable for $\beta < \frac{1}{2}$. 

The estimate of the anisotropy is still a matter of concern though: $\beta$ appears as $L^{-2 \beta}$. This term gives rise to a bimodality in velocity space, which cannot adequately fit the single-peaked velocity distribution observed for local halo members. Hence any anisotropy measurement with this model is condemned to fail, in most cases with an azimuthal bias caused by depopulating the velocity distribution around $V \sim 0$, which also explains why the anisotropy is at odds with observations of a strongly radially dominated velocity ellipsoid for local stars \citep[see e.g.][]{smi07,SAC}.

Equation (6) of D11 states that the mean rotational velocity for model (\ref{model_df}) depends on radius only. By computing 

\begin{equation}
\langle v_\phi \rangle =\frac{\int \textrm{d}^3 v \textrm{ } v_\phi F_{\eta,\beta}(E,L,L_z) }{\int \textrm{d}^3 v F_{\eta,\beta}(E,L,L_z) }
\end{equation}

\noindent we see that this is true only in the limit $\Delta \rightarrow 0$. Indeed, one can show that if $\tanh(L_z/\Delta) \neq {\rm sgn}(L_z) $, then $\langle v_{\phi} \rangle$ depends on the full 3D position. We will therefore adopt this assumption even though it was not stated in D11.

Furthermore, we note that the choice of the potential-density pair gives rise to an unphysical mean velocity in the inner Galaxy. According to eq. 6:

$$
\lim_{r \rightarrow 0}  | \langle v_\phi \rangle | =\infty.
$$

\subsection{Model-independent estimators of rotation}\label{sec:estimators}
We present here three model-independent estimators of rotation: one relies on $\vlos$ information only \citep[a similar approach has been previously adopted by][]{fre80} and the other two on the full 3D motion. SDSS proper-motion at these distances are of the size of the noise associated with them. All the same, given that the estimators rely on averages on the sky, valuable information can still be extracted from them provided that their errors are random and not systematic. To set ourselves in such a condition we correct for the known systematics of astrometric frame-dragging  \citep[][]{sch12}; if there is any other significant systematics left, the estimators will be biased differently and will therefore contradict each other (see SBA12). Further, one of them outputs the U-component of Sun velocity and the latter information can be directly compared to the values in literature to assess the presence of proper-motion systematics significant enough to bias the estimator.

\subsubsection{$\vlos$ estimator}

Inferring the rotational signature using only $\vlos$ presupposes knowledge of the global velocity distribution: i.e. how the velocity ellipsoid projects into $\vlos$. In fact, it is sufficient to assume that  the azimuthal velocity is not correlated with the other components (the velocity ellipsoid tilts between $U$ and $W$, but not $V$) and that the means of the other two components are $\sim 0$. This minimal set of assumption allows us to measure rotation from the de-projected $\vlos$ on the azimuthal direction:

\begin{equation}
\vlos = ({\bf v}-{\bf v}_\odot) \cdot {\bf \hat{s}} \Rightarrow {\bf v} \cdot  {\bf \hat{s}}= \vlos + {\bf v_\odot} \cdot {\bf \hat{s}},
\end{equation}
\noindent where $ {\bf \hat{s}}$ is the l.o.s. unit vector, ${\bf \hat{e}}_\phi$ is the unit vector in the azimuthal direction of Galactic cylindrical coordinates and ${\bf v_\odot}$ is the Sun's velocity w.r.t. the LSR. Therefore:

\begin{equation}\label{deproj}
\langle v_\phi \rangle \simeq  \Big\langle \frac{\vlos + {\bf v_\odot} \cdot {\bf \hat{s}}}{{\bf \hat{s}}\cdot {\bf \hat{e}}_\phi} \Big\rangle .
\end{equation}

Hereafter, we shall refer to the de-projected l.o.s. velocity as 
$$
\vdep:=\frac{\vlos + {\bf v_\odot} \cdot {\bf \hat{s}}}{{\bf \hat{s}}\cdot {\bf \hat{e}}_\phi}.
$$

\begin{figure}

\epsfig{file=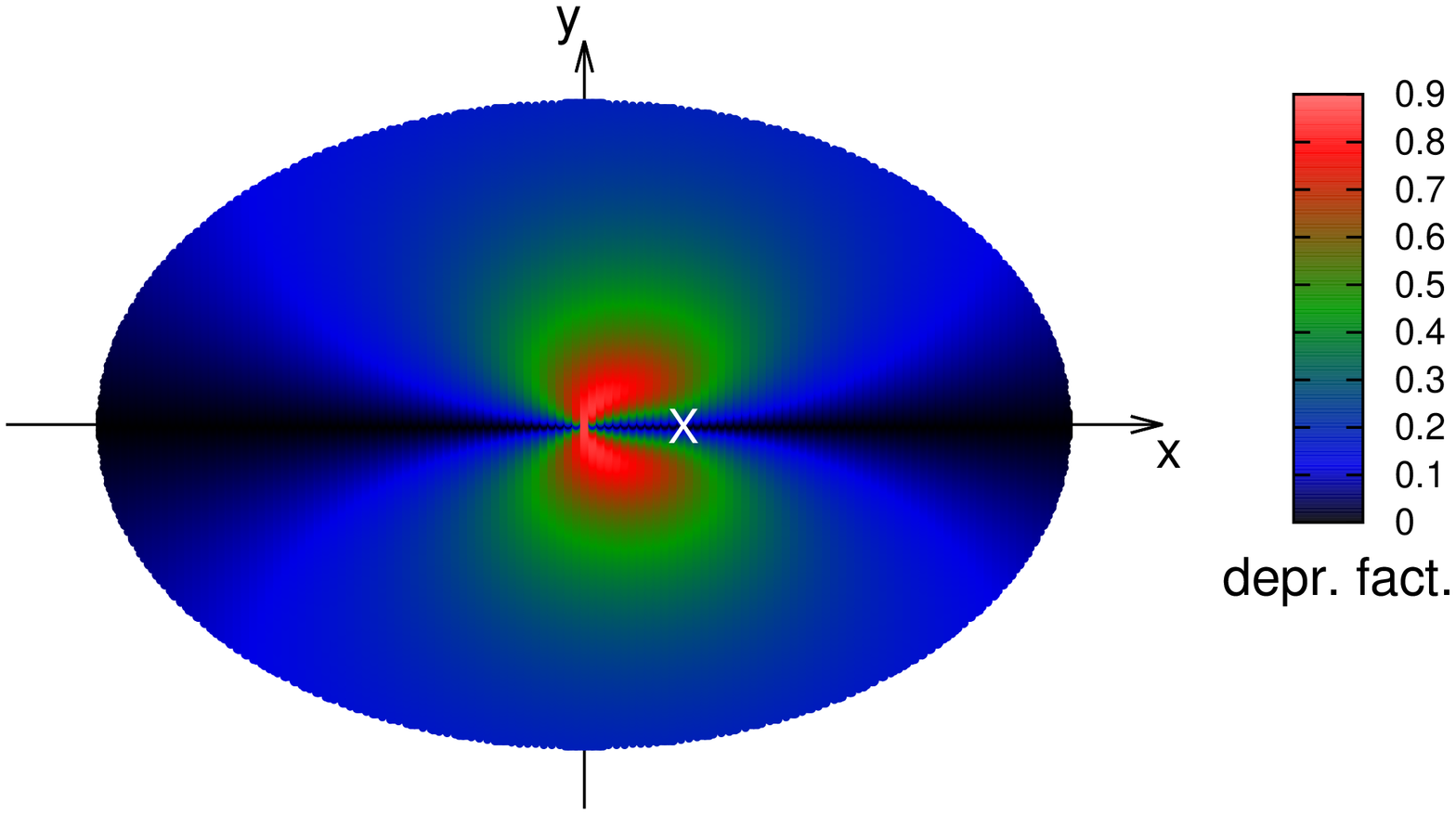,angle=-90,width=85mm}
\epsfig{file=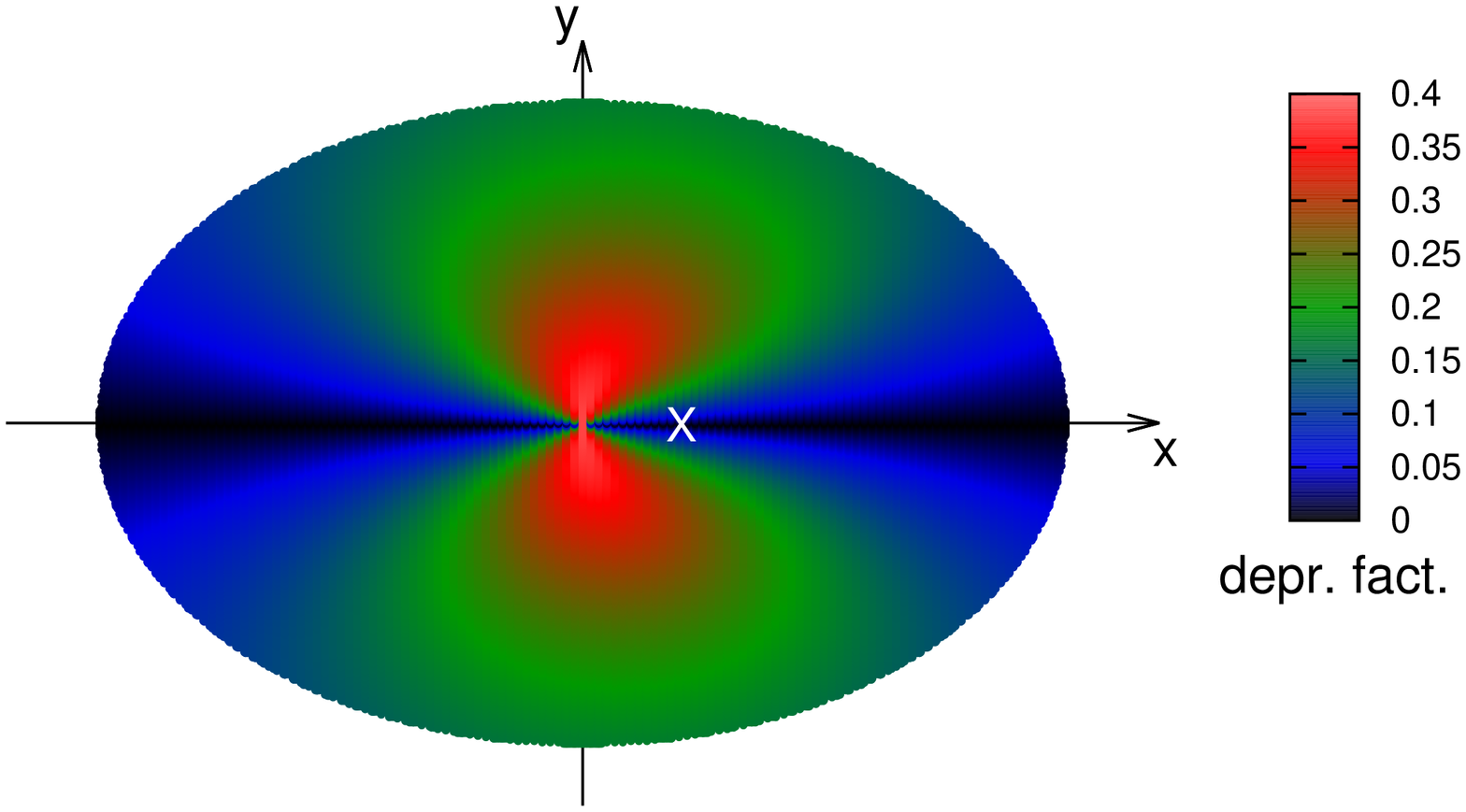,angle=-90,width=85mm}

\caption{Colour coded deprojection factor in the $(x,y)$-galactocentric Cartesian plane at altitudes of $|z| = 5 \kpc$ (top panel) and $|z| = 20 \kpc$ (bottom panel), limiting $\sqrt{x^2+y^2} < 45 \kpc$. The white cross denotes the Sun's position; note the different colour scales of the plots.}

\label{deprojfig}

\end{figure}

If $v_\phi$ is independent of position we can take a global average and expect that $\nabla_{\bf x} \langle v_\phi \rangle \sim 0$, where ${\bf x}$ is the 3D position vector. The best estimate of the mean streaming motion is then obtained by averaging the deprojected measurements scaled by their associated variances following a maximum likelihood argument. In practice, we fit a straight line in the  $x_j$, $\vdep$ plane with the ``weighted" least square routine of \verb"gnuplot" having marginalized over the other two position variables.  If $\vdep$ depends on position, the estimator still gives a best estimate of the position-dependent rotation velocity. The variances must account for the projection of the $U$, $W$ components of velocity along the l.o.s. as well as for the dispersion in the $V$ distribution $\sigma_V$, so that the final variance is:

\begin{equation}\label{weights}
{\sigma_\zeta}= \frac{\left( \sigma_{\vlos,\zeta}^2 + \sum_{j=1}^3 (\sigma_j \hat{\bf e}_j \cdot  {\bf \hat{s}})^2 \right)^{\frac{1}{2}}}{ {\bf \hat{s}}\cdot \hat{{\bf e}}_{\phi}},
\end{equation}

\noindent where $\sigma_{\vlos,\zeta}$ is the radial velocity measurement error for the $\zeta$-th star, and the sum describes the projection of the velocity ellipsoid into the l.o.s. with direction vectors $\hat{\bf e}_j$ and dispersion $\sigma_j$. 
The deprojection factor has a singularity, where ${\bf \hat{s}}$ is perpendicular to $\hat{\bf e}_{\phi}$. In Fig.\ref{deprojfig} we show ${\bf \hat{s}}\cdot \hat{{\bf e}}_{\phi}$ in the $(x,y)$-galactocentric Cartesian plane at altitudes of $|z| = 5$ and $|z| = 20 \kpc$. A deprojection factor of $0.1$ means that a rotation velocity of $20 \kms$ leaves a signal of just $2 \kms$ in the mean l.o.s. velocity. Hence the region where l.o.s. velocities can be reliably used is confined to small galactocentric radii and low altitudes with the best regions along the y-axis. The variation also implies that weighting the data points according to their uncertainty is mandatory to obtain any meaningful results. Neglect of $\sigma_\zeta$ will lead to large errors (see \S \ref{sec:geocheck}) and fluctuations of the results.\footnote{We suspect this is the case in study of \cite{hat}, who do not mention any care towards this issue.} 
Fortunately we do not require exact knowledge of the uncertainty. Since variations in the numerator of (\ref{weights}) are comparably small and since weighted least squares only improve the statistics but do not introduce any bias, we can safely set $\sum_{j=1}^3 (\sigma_j \hat{\bf e}_j \cdot \hat{s})^2$ constant, as long as we recover the approximate shape around the singularity.

\subsubsection{3D estimators}\label{sec:3Dest}

We have extracted rotational information out of $\vlos$ alone both via a model and via a direct estimator. $\vlos$ does not hold the full kinematics though and we could be missing stream-like structures in our sample with peculiar motions. Therefore to disentangle the possible bias by streams, we need to make use of the full 3D velocity. 

This implies that the major part of velocity information actually comes from proper motions. While velocities of individual stars are not reliable - the error is approximately as large as the signal - large number statistics solve that problem for aggregate quantities like the average rotation.\footnote{Even if the noise is actually bigger than the signal, which is the case for SDSS measurements of stars beyond $20 \kpc$ from the Sun, this introduces no preferred direction of motion and hence no bias.} The error in the mean quantities is anyway best estimated from the as-observed scatter and does not require a-priori knowledge of the proper motion (random) errors. 

More of a concern are systematic errors on the proper motions like astrometric ``frame-dragging". We checked that neither the corrections from \cite{sch12} nor differences between the data releases can alter our results, for details please see subsection \ref{sec:daterr}.

Furthermore, SBA12 show that the mean streaming motion can be recovered from the position-dependent component of velocity ($U$) in the local frame:

\begin{equation}\label{proj}
U={\bf v} \cdot \hat{\bf e}_x =v_R \cos \alpha  + v_\phi \sin \alpha,
\end{equation}

\noindent where the velocity vector of the star ${\bf v}$ is projected by $\hat{\bf e}_x$ onto the local radial direction, $v_\phi$ is the effective rotation and 

\begin{equation}
\alpha=\arctan \left( \frac{d \sin l \cos b}{R_0 - d \cos l \cos b}\right)
\end{equation}

\noindent is the angle between the projection onto the plane of the long axis of the velocity ellipsoid and the Sun-GC line (see equations 26-28 in SBA12). The average value of $U$ depends on $\sin \alpha$, so that we can estimate the global mean $v_\phi$ from the linear regression:

\begin{equation}
\langle U \rangle =\langle v_\phi \rangle \sin \alpha - U_0.
\end{equation}

The linear fit recovers both the mean rotation and the $U$-component of Sun's velocity, $U_0$. 

Finally the direct estimate of the mean streaming velocity of the sample can serve as consistency check that the $U$-estimator is not biased from proper-motion systematics: if it were, contradictory results would emerge between the two estimators given their different bias.

\subsection{Sample systematics}

A broken kinematic signature across different sub-divisions of a sample is the consequence of either a distance error, a pipeline error or a stream-like structure. Provided that tracers of all prominent unrelaxed sub-structures have been removed from the sample, no matter how we sub-divide our sample, each sub-set must have the same kinematics. Whatever portion of the sky we look at, we have to obtain the same rotation estimate, 
within the uncertainty. In particular, splitting the sample in random sub-samples decreases the statistical signal-to-noise and consequently enhances the allowance for discrepancy in the signature between two bins. Therefore, an anomaly among different sub-divisions of the sample is even more concerning and serves as a strong indication for a bias in the analysis.

\subsubsection{Distance errors}\label{sec:geocheck}  When we recover the rotational signature of a population of stars where only $\vlos$ is available or reliable, an error in distance induces changes in both the angle between the line of sight and the vector from the star to the Galactic Centre, and the projection of $\vlos$ onto the plane tangential to the direction to the Galactic centre. Thus a distance error influences directly the rotational signature. This will not be a constant systematic, but a function of the position on the sky. Distance errors shift stars en masse in velocity space. Distance over-estimates tend
to shift velocities upwards towards the escape velocity, thus narrowing the range of physically accessible velocities to be considered. An increase in velocity will nearly always reduce the likelihood of a star when a model is being used to calculate the probability of the data; by carrying a star beyond the escape velocity they can even reduce the star's probability to zero. When a direct estimator is used, a distance over-estimate tends to increase the computational weight of a given star.

\begin{figure}

\epsfig{file=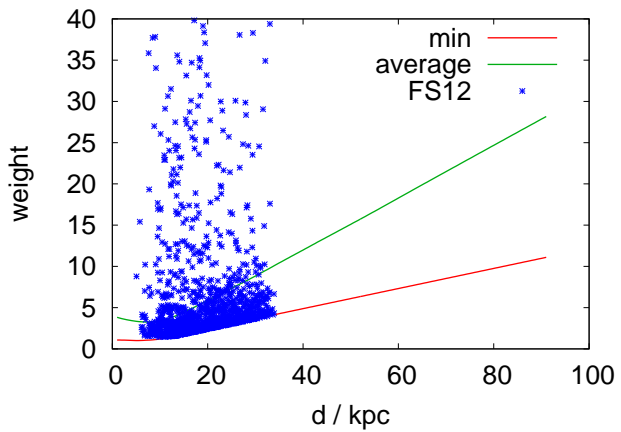,angle=0,width=85mm}

\caption{We move each star in the FS12 sample along the l.o.s. to put it at distance $d$, then calculate $\langle 1/(\hat{\bf s} \cdot \hat{\bf e}_\phi) \rangle$ and plot it versus $d$ (green line). The red line represents the minimum $1/(\hat{\bf s} \cdot \hat{\bf e}_\phi)$: both are almost perfectly linear functions of distance beyond 10 kpc. Blue crosses depict the stars from FS12 at their actual heliocentric distance. For graphical convenience we do not show the $6 \%$ of stars which exceed the vertical range. For better comparison (see \S 5) with the analysis of Deason et al. (2011), we extend the distance range to $90 \kpc$.}

\label{weight_scale}

\end{figure}

This effect does not change the sign of the rotation contribution of individual stars, but it distorts the values with a significance that scales with distance from the observer. In Fig. \ref{weight_scale} we see that the factor by which the star's velocity is multiplied 
increases almost monotonically heliocentric distance. 
Indeed, distant stars are intrinsically a bad tracer if proper-motion is not available: at large radii the streaming motion $v_\phi$ contributes very little to $\vlos$; in fact $\vlos + {\bf v_\odot} \cdot {\bf \hat{s}} \approx v_r$ and $v_\phi$ is dominated by the tangential component of velocity (hereafter $\vmu$). In order to change $\vlos$ by a small amount, $v_\phi$ needs to change significantly: the signal-to-noise scales inversely with radius making rotation estimated via $\vlos$ of distant stars, dominated by noise.

\subsubsection{Stream-like structures} 
A discontinuous rotational signature can hint at the presence of a sub-structure in the data: a stream for example can lead to inconsistencies between different rotation estimators, esp. for the $v_{los}$ based estimator and model.  Even though we do not hunt for substructures here, we must allow for unidentified accretion remnants mimicking rotation. To break the degeneracy between biased data and a real stream, we need to identify what is characteristic of a substructure and what is not. An unrelaxed substructure is identified as such by its peculiar signature in parameter space (both physical and kinematic). Correlations between kinematics and metallicity or between colour and kinematics are examples of signatures of a possible accretion event. By contrast binning in luminosity type acts like a random selection w.r.t. kinematics and hence a coherent signature across the bins must be observed. Further, provided that all detectable over-densities in observable space have been removed, binning the sample spatially implies that even if there are remnants of accretion events left they are washed out by the smooth component and thus a coherent signature is expected across all the sub-samples.

\subsubsection{Pipeline Systematics} 
A bias in the template fitting of the pipeline can break the rotational signature as we move from the left ($l<180$) to the right ($l>180$) side of the Galaxy. The spatial dependence will make a genuine coherent streaming motion stronger on one side and faint, or even of opposite sign, on the other side. In general, a higher degree of caution must be adopted especially for hot metal-poor stars, since their spectra are almost free of metal-lines and a line-of-sight velocity determination based on the very broad Balmer lines is vulnerable to template misfits.

\section[]{Sample selection and calibration}

To study the Galactic halo in situ we need tracers that are bright enough to allow for distances $> 5 \, \kpc$. The studied stars should have relatively well-determined absolute magnitudes and should not be easily mixed up with other types of stars, e.g. foreground dwarfs from the Galactic disc(s). Late type giants are problematic by their easy confusion with dwarf stars and by their steep relation between colours and magnitude, which also amplifies reddening uncertainties. The optimal objects would be RR Lyrae stars \citep[cf.][]{kl1}, but in addition to spectroscopy they require a good coverage in time to resolve their oscillations and hence are not globally available \citep[][]{Sesar10}.

Blue horizontal branch (BHB) stars have the advantage of a relatively narrow absolute visible magnitude distribution that shows only weak dependence on metallicity. It should be kept in mind, however, that there is the blue hook with fainter objects on the hot side of the BHB, and that, despite most Galactic stars being redder in colour, they mingle in observations with the fainter blue stragglers and main-sequence objects. Three techniques have been developed to identify BHB stars by looking at hydrogen lines: measuring the size \citep[][]{phi94}) or the steepness \citep[][]{Chalonge73} of the Balmer Jump and using the width of the Balmer lines \citep[][]{sea66}. After identification BHB stars are treated as standard candles. It should be kept in mind, however, that the use of BHB stars biases any sample to old, metal-poor populations. Moreover, theoretical models show a metallicity dependence of order $\sim 0.3 \mag$ in the V-band magnitude (and similarly in the Johnson g-magnitude), emphasizing the need to control any result not only for contamination, but also for absolute magnitude uncertainties \citep[][]{P1}.

We construct two samples from SDSS DR9: one selected via filters in colour, surface gravity and effective temperature and another selected by X11 via direct use of Balmer lines. We will refer to the first sample as photometrically selected calling it FS12 and to the second sample as spectroscopically selected, calling it X11. Even though the FS12 sample is drawn from data for which spectra are in fact available, the chosen labels aim at emphasizing the difference between using and not-using Balmer-line shape measurements in the selection process (see \S \ref{sec:X11}). To both samples we apply geometric cuts to exclude the disk and Sagittarius.

Astrometric accuracy is of crucial interest in our study: the statistical error on SDSS proper-motions is of the order of $\sim 3 \, {\rm mas}{\rm yr}^{-1}$ \citep[][]{Munn04,mun04err,don11}. 
We constructed our statistics (see \S \ref{sec:3Dest}) such that we do not require any knowledge on the proper-motion error distribution: the only implied assumption in our error estimates is that the observed velocity distributions including the errors follow a roughly Gaussian distribution. The violation of this assumption does not affect our parameter estimates, but leads to a mild misjudgement of their errors. Proper-motion systematics are more of a concern: we assess their effect in our analysis in detail in \S \ref{sec:daterr}.

\subsection{Selection via photometry}

We construct a sample of BHB stars, drawn from SDSS \citep[DR9][]{ahn12}\footnote{See Appendix B for the SQL query.}. The selection benefits from the extensive work done 
by \cite{yan00} and \cite{sir04}, who revised the BHB identification procedure working with SDSS data.

Both agree on an initial colour filter such as: $g-r \in [-0.4,0]$ and $u-g \in [0.8,1.4]$. The comparison with ultra-violet colour $u-g$ provides indirect information on the Balmer jump and hence is an indirect gravity estimator. Still in this colour regime (which we will denote $\mathcal{R}$) there are contaminations from other A-type stars (mostly blue stragglers) and from F stars due to ``intrinsic variations and photometric errors'' \citep[][]{sir04}. While F stars can be filtered out by their lower temperature, A-type stars mostly differ by their surface gravity, so that both groups consequently use the Balmer line width to separate them. 
Their understanding of the problem can be summarized as follows: main-sequence A stars can be distinguished from lower surface-gravity BHB stars (which lie in the region $\log g \in [2.5,3.2]$) by looking at the width of the $H\delta$ lines alone  \citep[][]{yan00} or $H\delta$ and $H\gamma$ lines \citep[][]{sir04}. The discrimination is not sharp and one must still expect  contamination of order $\sim 10 \%$ \citep[][]{sir04}. The situation deteriorates towards fainter magnitudes and hence noisier spectra:  \cite{sir04} estimate contamination of order $\sim 25 \%$ at $g>18$. Indeed \cite{lee08} show that the low spectral resolution in SDSS and limited signal to noise lead to larger uncertainties in stellar parameters including $\log(g)$ up to a level where subgiants and dwarfs get mixed up. This represents a serious challenge for the proposition that stars can be cleanly sorted in type using only the $\log(g)$ estimates from the SEGUE stellar pipeline \citep[e.g.][]{car07,car10,de11}.

\begin{figure}
\epsfig{file=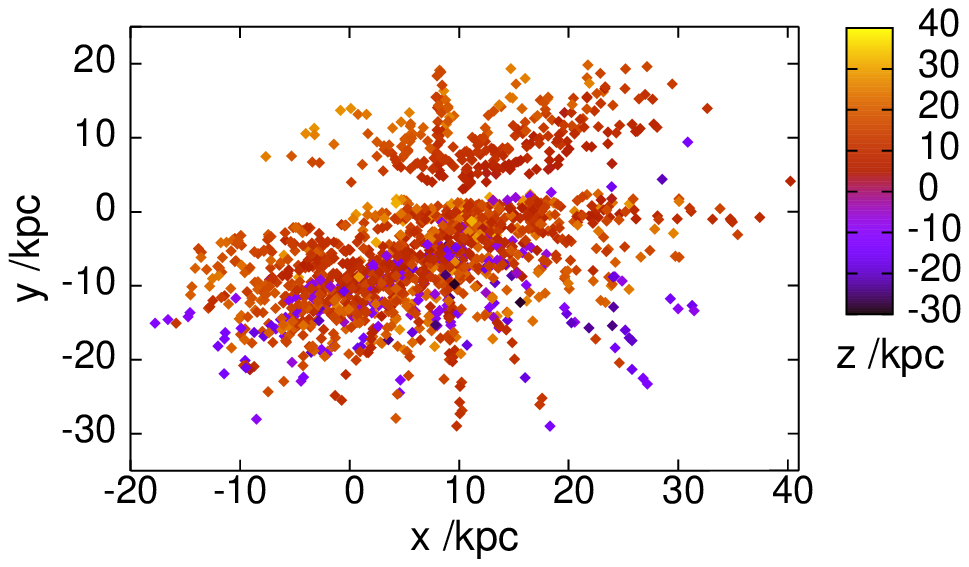,angle=0,width=85mm,height=40mm}
\epsfig{file=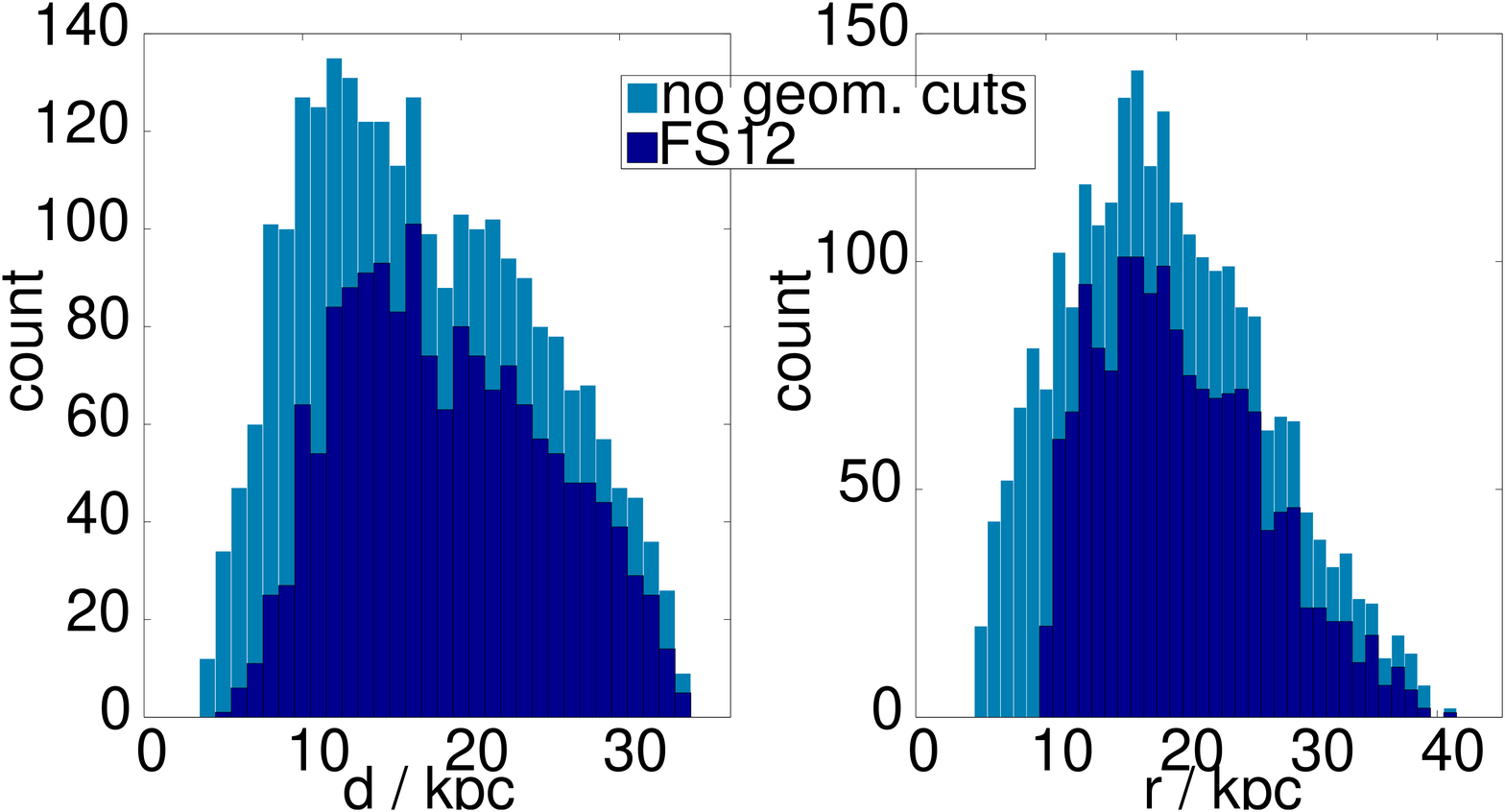,angle=0,width=85mm,height=40mm}

\caption{Geometry of the FS12 sample (top panel): distribution in the $(x,y)$-plane with the origin at the Galactic Centre, altitude z is colour-coded (Cartesian Galactocentric reference frame). $207$ out of $1585$ stars ($13 \%$) are in the southern galactic hemisphere. The bottom panels show the heliocentric (left) and Galactocentric (right) distance distribution. The dark blue histograms refer to the FS12 selection (\ref{FScuts}), while the light blue ones show the sample without the geometric cuts.}

\label{FS12geo}

\end{figure}

According to the dangers reviewed in the previous paragraph, we minimise the risk of contamination from other luminosity types, by selecting:

\begin{equation}\label{FScuts}
\left\{ \begin{array}{l}
g<18\\
2<\log g<3.3\\
0.8<(u-g)_0<1.4\\
-0.4<(g-r)_0<0 \\
7250<T_{\rm eff}/{\rm K}<9700\\
|z|>4 \kpc \textrm{ } \& \textrm{ } r>10\kpc\\
(\alpha,\delta) \notin {\rm Sgr}
\end{array}\right.
\end{equation}

\noindent  where the last two criteria are geometric cuts to exclude disk stars and members of the Sagittarius stream (e.g. D11)\footnote{The masking of Sgr is performed in right ascension-declination space ($\alpha, \delta$) according to the polygon kindly provided by the authors of D11 on request (see Appendix B).}. The estimates of surface gravity, effective temperature and metallicity are from the analysis of \cite{wil99} as it was specifically designed for hot stars. This leads to a sample of 1585 stars. In Fig. \ref{FS12geo} we plot the geometry of the sample: the top panel shows the distribution in the $(x,y)$-plane colour-coded according to the z-coordinate (Cartesian Galactocentric reference frame) and in the bottom panels we show the heliocentric (left) and galactocentric (right) distance distributions.

\begin{figure}

\epsfig{file=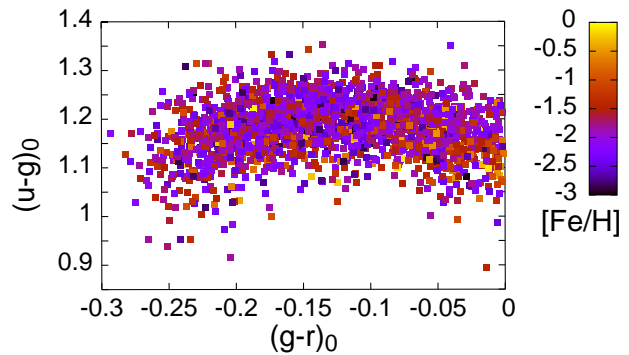,angle=0,width=85mm}

\caption{BHB stars drawn from SDSS DR9 included by the selection (\ref{FScuts}) in the $(u-g)_0$, $(g-r)_0$ plane. Every point is coloured according to the metallicity of each star.The commonly adopted colour cuts appear to include some part of the \emph{blue hook}.}

\label{ccplane}

\end{figure}

We further investigate the purity of our sample in the colour-colour plane and find grounds to suspect the presence of the \emph{blue hook} in the stripe $-0.3<(g-r)_0<-0.2$ (see Fig. \ref{ccplane}) which suggests that a $u-g$ dependent cut in the $g-r$ colour might more cleanly identify the horizontal branch. To a zero-th order approximation, a constant cut at $(u-g)_0> 1.15$ (rather than $(u-g)_0>0.8$) is cost-effective and will therefore be our initial choice to calibrate the reliability of (\ref{FScuts}). The tighter cut in $(u-g)_0$ shrinks the sample to 1297 objects. 

Even though our stringent selection criteria cause a drop in the sample size w.r.t. previous works (e.g. the sample of D11 has $\sim 3500$ stars due to more generous filters in surface gravity and apparent magnitude), none of our filters correlate with kinematics and thus they do not prejudice our analysis. They also do not bias the distributions of physical parameters such as for example metallicity: the distribution peaks at $\fe=-1.9$, in good agreement with the estimates from the samples of \cite{x08,x11}, who do not impose any cut in surface gravity or apparent magnitude for example. Further, we judge the increase in the statistical noise in the model fits a convenient price to pay in light of the reduced risk of contamination. The fact that our selection criteria in $\log g$ and $g$-band magnitude do not bias the inferred kinematics is confirmed by a second sample selected via more generous photometric filters, but combined with Balmer-line shape measurements, which presents an homologous rotational signature (see following Section).

\subsection{Spectroscopic selection}\label{sec:X11}

X11 presented a spectroscopically selected sample of $\sim 5000$ BHB stars drawn from SDSS DR8 and which has been kindly provided to us. The advantage w.r.t. a selection via photometric information is that the three independent methods of identification separate BHB stars from blue stragglers and main-sequence stars more cleanly by combining colour cuts with Balmer-line shape measurements: the scale width-shape method fits Sersic profiles to the $H\gamma$ lines \citep[cf.][]{cle02,sir04,x08,x11}. The parameters that describe such a profile are essentially the amplitude ($a_\gamma$), the dispersion ($b_\gamma$) and the steepness ($c_\gamma$) of the modified Gaussian that is fitted to each $H_\gamma$ line:

\begin{equation}\label{Sersic}
y(\lambda)=1.0-a_\gamma \exp \left[  - \left( \frac{|\lambda - \lambda_0|}{b_\gamma} \right)^{c_\gamma} \right].
\end{equation}

\begin{figure}
\epsfig{file=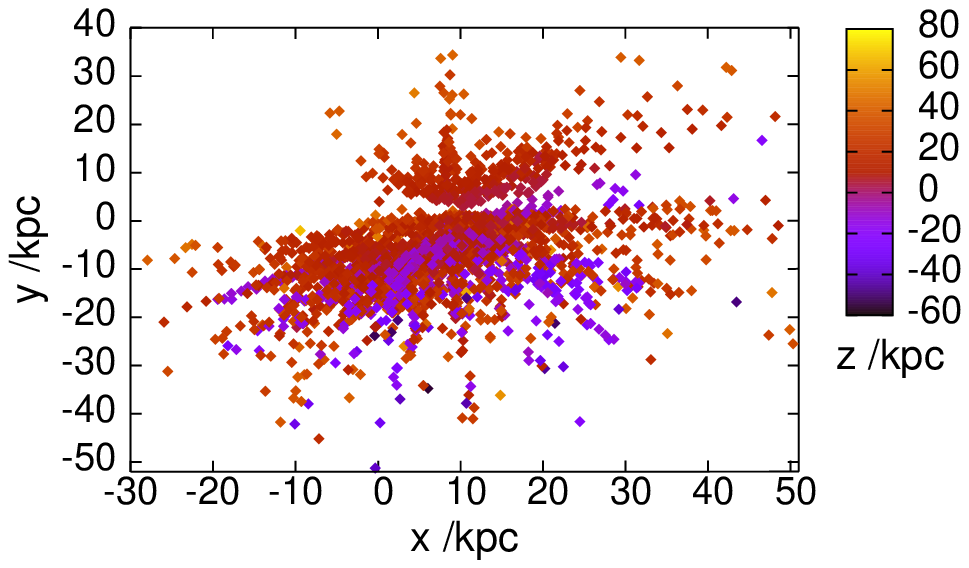,angle=0,width=85mm,height=40mm}
\epsfig{file=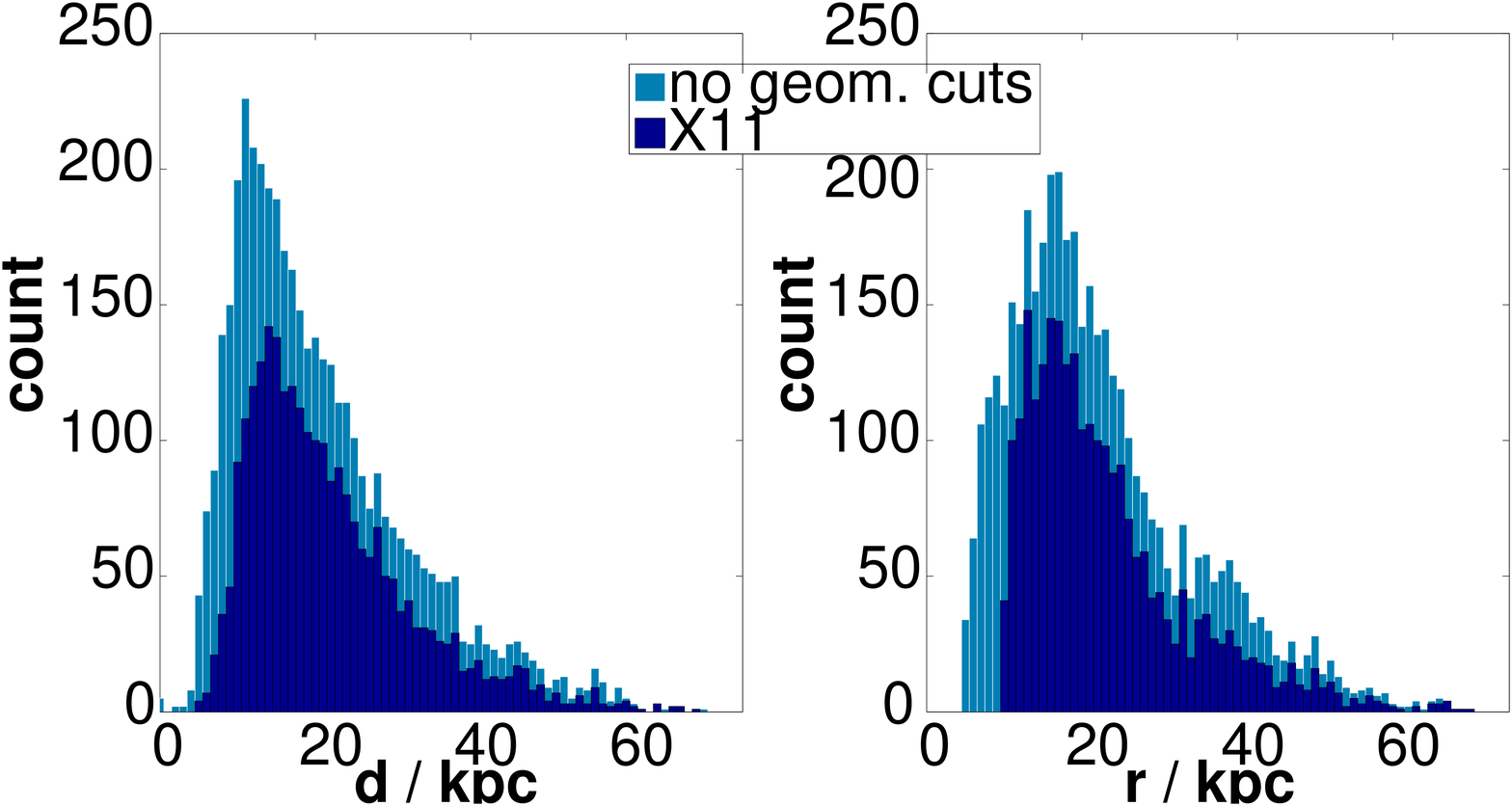,angle=0,width=85mm,height=40mm}

\caption{Same as Fig \ref{FS12geo}, but for the X11 sample. Here $526$ out of $2563$ stars ($21 \%$) belong to the souther galactic hemisphere.}

\label{X11geo}

\end{figure}

We retrieve X11's stars in DR9 and use the astrometry of the latest data release; we exclude contributions from the disk and the region of Sagittarius by applying the same geometric criteria of the previous section: this leads to a sample of 2563 stars (2455 with $\fe \in [-3,0]$).  We show its geometry in Fig. \ref{X11geo}. Among these $\sim 1400$ have $c_\gamma \ge 0.95$, which is the limit value for which the method of \cite{cle02} has been calibrated. Given that neither \cite{sir04} nor \cite{x11} have re-calibrated the method in the region $c_\gamma  \ge 0.95$ we cautiously flag these stars and first consider them in our analysis but later analyse them separately.

We retrieve 1400 stars of our photometrically selected sample in the sample of X11, which implies that almost $90 \%$ of the FS12 sample is included in X11: the latter sample was selected using a classification based on both Balmer-lines and colour cuts and therefore constitutes the more robust identification of BHB stars to the current understanding. Consequently, cross-matching our FS12 sample with the sample of \cite{x11} provides a rough estimate for the level of contamination from non-BHB stars that we can expect: i.e. $\sim 10 \%$, which very well matches the independent prediction of \cite{sir04}. Note that when we compare the sample of D11 with X11, only $47 \%$ of their stars have Balmer-line shapes identifiable with BHB type stars, even though D11 drew their sample from DR7 and X11 is based on DR8, which is significantly larger. If we exclude high surface gravity and faint stars from the sample of D11 in DR7, the proportion of stars matched by \cite{x11} increases to $71 \%$, meaning that the majority of the stars we rejected are non-BHB stars according to \cite{x11}.\footnote{Note that the value reported in D11 at the beginning of their \S 3.1 in these regards can be somehow misleading. The authors report that $88 \%$ of the stars of \cite{x08} (based on DR6) is found in their larger sample. But this actually means that only $41 \%$ of the stars in D11 sample is found in the one of \cite{x08} and we do not have any reassurance on the level of contamination for the remaining $59 \%$ of the stars in their sample of ($\sim 3274$ objects).}

\subsection{Absolute Magnitude}\label{sec:absmag}

The second issue that needs careful handling is the estimate of the absolute magnitude for the BHB stars. \cite{sir04} show that the assumption of a constant luminosity and mass, temperature-independent for $T_{\rm eff} \in  [8000,12000] {\rm K}$, \citep[][]{bae01} would lead to an average overestimate in distance modulus of $0.18$ mag with respect to a temperature-dependent relationship for the luminosity \citep[][]{dor93}. Furthermore, the popular value of $M_g=0.7$ was calculated by \cite{lay96} for halo RR Lyrae stars and then adopted for BHB stars given the similarity in their absolute magnitude. The justification of this approximation depends on the blue response of the filter in use \citep[][]{yan00}. \cite{pres91} report that BHB stars in the blue end of $\mathcal{R}$ are $0.7$ mag fainter in $g$ than the RR Lyrae stars\footnote{Both \cite{lay96} and \cite{pres91} report results for the $V$ magnitude, which \cite{yan00} usefully converts into the $g$ magnitude via the colour transformation of \cite{fuk96}.}. Therefore the uncertainty reported for the absolute magnitude of BHB stars needs to be interpreted generously given the non-standardization of the blue filters and the consequent difficulty in comparing estimates from different surveys.

\cite{P1} show that the assumption of a constant $M_g$ for BHB stars introduces systematics correlated with colour and metallicity and propose a colour and metallicity dependent distance calibration for BHB field halo stars:

\begin{eqnarray}\label{Mg}
M_g (\,(g-r)_0,&\fe) &=  0.0075 \, \exp(-14.0 {(g-r)_0}) + \nonumber \\
& & + 0.04\left(\fe+3.5\right)^2+ 0.25.  
\end{eqnarray}

\noindent This formula covers the theoretical expectations from the BASTI isochrones \citep[][]{pie04,pie06}, and in statistical tests on field stars proves to be more accurate than other available calibrations \citep[e.g.][]{sir04,de11_416}, which were only based on globular clusters. In particular, (\ref{Mg}) hits the precision threshold allowed by the sample size, on FS12 the average fractional distance error is $0.01 \pm 0.04$ and $0.02 \pm 0.03$  on X11, and is the only calibration to pass the ``falling sky" test of SBA12 \citep[][\S 4.1 and 5.6 respectively]{P1}. Crucially in this context, it covers the metallicity dependence and hence avoids overestimating the distance to metal-rich objects and underestimating that to metal-poor ones resulting in metallicity-dependent kinematic biases. We will adopt the above formula throughout the paper, unless otherwise stated or when comparing to D11, who assume a constant BHB absolute magnitude.

\section{Rotational Signature of the Milky Way stellar halo}

We present below four independent estimates of the rotation of the Milky Way stellar halo, traced with the two different samples of BHB stars presented in the previous section.

\begin{table*}
	\centering
	\begin{tabular}{p{0.30\linewidth}p{0.22\linewidth}p{0.22\linewidth}}
	\hline
 \hspace{2.5cm} {\rm all} & $\fe \in [-3,-2]$& $\fe \in [-2,0]$ \\ \hline
{\bf FS12 } \hspace{2.5cm} & & \\
$\begin{array}{lc} \eta & 0.83 \pm 0.08 \\ \langle \pi^{-1} (\vlos) \rangle  &  17 \pm 9 \kms  \\ \langle v_\phi \rangle_{U{\rm-est}} & 0 \pm 11 \kms \\ \langle v_\phi \rangle & 6 \pm 6 \kms \end{array} $& $\begin{array}{c}  0.84 \pm 0.17 \\ 11 \pm 15 \kms \\ 2 \pm 17 \kms \\ -3 \pm 10 \kms \end{array} $& $\begin{array}{c} 0.79 \pm 0.12 \\ 22 \pm 12 \kms \\ -1 \pm 14 \kms \\ 13 \pm 8 \kms \end{array} $\\ \hline
{\bf X11 } \hspace{2.5cm}  & & \\ 
$\begin{array}{lc} \eta & 1.04 \pm 0.08 \\ \langle \pi^{-1} (\vlos) \rangle  & -3 \pm 7 \kms \\ \langle v_\phi \rangle_{U{\rm-est}} & -13 \pm 10 \kms \\ \langle v_\phi \rangle & -7 \pm 5 \kms \end{array}$ & $\begin{array}{c}  1.22 \pm 0.15 \\ -18 \pm 11 \kms \\ -7 \pm 16\kms \\ -4 \pm 8 \kms \end{array}$& $\begin{array}{c} 0.90 \pm 0.10 \\ 9 \pm 9 \kms \\ -18 \pm 13 \kms \\ -9 \pm 7 \kms \end{array} $\\
	\hline
	\end{tabular}
	\caption{Estimates of rotation using both $\vlos$ information only and full 3D motion on two samples drawn from SDSS DR9. $\eta$ is the rotation parameter from model (\ref{model_df}) and implies prograde rotation for $\eta<1$ and retrograde motion for $\eta>1$. $\langle \pi^{-1}(\vlos) \rangle$ is the rotation estimate from the $\vlos$-estimator and $\langle v_\phi \rangle_{U-est}$ the one from the $U$-estimator (see \S 2.2).}
\label{res_sum}
\end{table*}

\begin{figure}

\epsfig{file=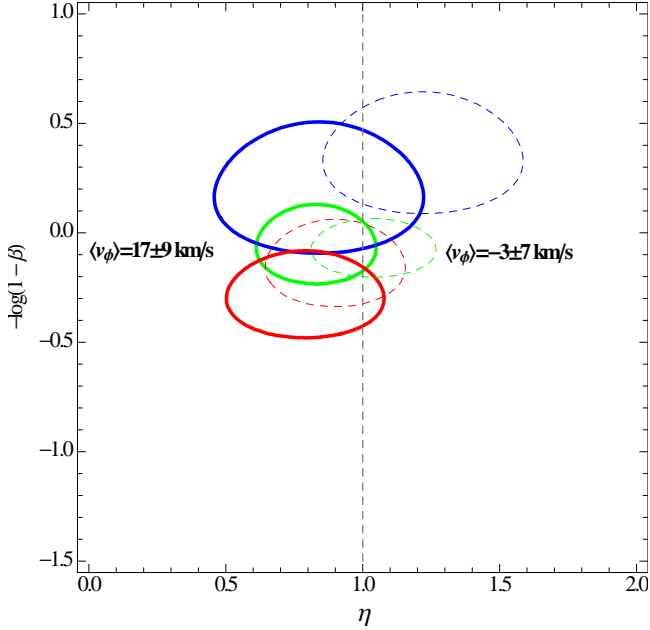,angle=0,width=85mm}

\caption{ Rotation and anisotropy of BHB stars from SDSS DR9 estimated via likelihood of the model (\ref{model_df}) given the data. Green line is the one sigma contour for the full sample, red line is associated with the metal rich stars and the blue one with the most metal-poor objects. Full lines are associated to the sample selected via (\ref{FScuts}), dashed lines refer to the spectroscopically selected sample from X11. The model-independent $\vlos$ estimator (see \S 2.2) agrees with the findings of the model. Both methods use $\vlos$ information only.}

\label{model_results}

\end{figure}

The likelihood analysis yields a very weak rotation on both samples, consistent with non-rotation. On FS12 we do not observe any significant trend in the rotational parameter $\eta$ (see \S \ref{sec:model}) across metallicity: $\eta_{\fe<-2}-\eta_{\fe>-2}=0.05 \pm 0.21$, while there is a gradient in the X11 sample:  $\eta_{\fe<-2}-\eta_{\fe>-2}=0.32 \pm 0.18$. This mismatch disappears on the sub-sample where FS12 and X11 overlap ($\sim 1400$ stars). The $\vlos$ estimator confirms all the above conclusions. Fig. \ref{model_results} presents these results together with the anisotropy estimates from the model in the different cases: FS12 favours an anisotropy consistent with zero, while X11 suggests a very weak radial bias: $\beta = 0.08 \pm 0.05$. It is worth remarking that as pointed out in \S 2.1.2, the model is intrinsically inadequate to estimate the anisotropy due to the $L^{-2 \beta}$ term - therefore, the obvious discrepancy to the radial bias in local samples with full 3D information is just a sign for its failure. We note that adopting either a constant distance scale ($M_g=0.7$, to be consistent with D11) or the colour-metallicity dependent calibration (\ref{Mg}) does not alter the above conclusions.

When we apply the $U$ and $v_\phi$-3D estimators, they converge to the same conclusion of a rotational signal consistent with zero. On one side, the fact that two 3D estimators agree implies that whatever the nature of the errors affecting the proper-motion, they are either random or not significant enough to bias the average the estimators rely on. Indeed, were there still significant systematics, the estimators would diverge. Further, the $U$-estimator provides also the $U$-component of the Solar velocity. On FS12 we measure $U_0=15.7 \pm 7.4 \kms$ and $U_0=7.8 \pm 6.8 \kms$ on X11: within the error bars both agree with the accepted value of $U_0$, again supporting the reliability of the 3D estimators. All estimates are obtained applying the proper-motion correction by \cite{sch12}: in \S \ref{sec:daterr}, we quantify the effect of neglecting this amendment.

On FS12 all estimators agree in measuring no or very weak rotation in both metallicity subsamples (see Table \ref{res_sum}). However, on the X11 sample the 1D-estimators measure a slight, though significant, discrepancy between the kinematics of metal-poor and metal-rich stars\footnote{We call metal-poor stars the ones with $\fe \in [-3,-2]$ and metal-rich the ones with $\fe \in [-2,0]$. The ratio is 9:20 for both samples.}.

For the entire halo population, all estimators measure a rotational signature consistent with zero on both samples. On FS12, both the 1D and the 3D estimators agree that there is no significant discrepancy in the rotation of metal-poor objects w.r.t. metal-rich ones. On X11 the situation is less clear as the 1D estimators support the existence of a gradient in rotation across metallicity bins, while the 3D estimators disprove it.\footnote{Note though that an aberration in the l.o.s. velocity would shift the estimate of the 1D estimator leaving nearly unaffected the ones of the 3D estimators (see \S \ref{sec:daterr}). An aberration of $\pm 5 \, (10) \kms$ (following what measured by \cite{sch12} for blue metal poor stars, though for a different portion of the sky) will shift the $\pi^{-1}(\vlos)$-estimate of $-18 \pm 11 \kms$  by $\pm 9.7 \, (19.4) \kms$ while moving the 3D-estimator by only 1 (2) $\kms $.} This issue arises on the sample where we had already flagged $50\%$ of the stars as potentially problematic: in \S 6 we critique this discrepancy in detail.

\subsection{Notes on uncertainties}

\subsubsection{Uncertainties from datasets}\label{sec:daterr}

Proper motions of stars have significant systematic errors, which may affect the $3D$-estimators. The main systematic that affect the proper motions are ``frame-dragging" and chromatic aberration \citep[][]{kac09}. The astrometric frame in Sloan is fixed using galaxies, and can be ``dragged" by the inclusion of stars in the ``galaxy" sample \citep[see discussion in][]{sch12}. The correction developed by \cite{sch12} is designed for usual stellar colours and the chromatic aberration for the bluer BHB stars may be different. However, the \cite{Schneider10} quasar sample harbours only very few objects with observed $g-r < 0$ (about $10\%$) providing us with a meagre basis for tests. While for $g-r > 0$ the sample displays almost perfect stability, the blue end deviates at the $1-2 \sigma$ level towards smaller corrections. Fortunately, the correction on the DR9 sample shifts the average rotational velocity $\langle v_\phi \rangle$ by a moderate $\sim 0.6-1.6 \kms$ for the $\langle v_\phi \rangle$-estimator and by $\sim 1.1-1.6 \kms$ for the $U$-estimator towards retrograde motion.  Anyway, in most cases we are only interested in the rotation difference between two sub-samples, which is nearly unaffected. Therefore, even with the proper-motion uncertainty the mean $v_\phi$ provides valuable information to check the robustness of the model-dependent rotation estimates.

The heuristic exercise of perturbing our $\vlos$ by a systematic $\pm 5 \kms$ shows that the 3D-estimators are rather robust against systematics in the l.o.s. velocity:
\begin{equation}
\langle v_\phi \rangle_{\vlos \pm 5 \kms} - \langle v_\phi \rangle_{\vlos} \sim \pm 1 \kms.
\end{equation}
The 1D $\vlos$-estimator is on the contrary affected by a shift larger than the actual systematic introduced: 
\begin{equation}
\langle \pi^{-1}(\vlos) \rangle_{\vlos \pm 5 \kms} - \langle \pi^{-1}(\vlos) \rangle_{\vlos} \sim \pm 9 \kms.
\end{equation}
This strong deviation might be surprising at the first glance, since the left and the right side of the sky should balance each other. However, the sample is strongly asymmetric, and worse, the geometric factors in the 1D-estimator deprojecting the observed mean motion into a rotational signal amplify any systematic on the line-of-sight velocities. 
This confirms the necessity of validating rotational estimates via 3D-estimators and supports the conclusions of the latter whenever they disagree with the 1D-estimators.\footnote{ For example the kinematic trend across different metallicity bins found in the X11 sample and supported by the 1D-estimators only will in fact result in being most probably an artefact (see \S 6).}

As a further check we repeated all experiments on earlier data releases DR7 and DR8, before SDSS underwent significant changes especially to the astrometry \citep[][]{SloanDR6,SloanDR7,SloanDR8,Sloanerr,Munn04} and all the above statements are confirmed well within one sigma. For those previous data releases the selection via photometry was obtained with a more generous filter in surface gravity: $\log g \in [2.0,3.5]$, accounting for the higher surface gravity estimates from the SSPP by on average $\sim 0.2$ dex in DR8 w.r.t. DR9 \citep[][]{ahn12}. The care for the high surface gravity stars is justified by the risk of contamination from other A-type stars when $\log g> 3.2$ \citep[e.g. blue stragglers,][]{sir04}. All the same, applying the more generous filter $\log g \in [2.0,3.5]$ also to the FS12 DR9 sample does not alter our results: the rotation measured by the model shifts by $\sim 5 \%$ and the estimators by $1-5 \kms$, but not in a systematic way. With this cut the sample is enlarged by $\sim 200-400$ stars for $(u-g)_0>1.15$ and $(u-g)_0>0.8$ respectively.

\subsubsection{Systematics in the analysis}

We repeat the same analysis applying the more generous cut in $u-g$ colour discussed in \S \ref{sec:X11} to include what we identified with the onset of the blue hook in the colour-colour plane. For our sample we find that the effect is negligible: the discrepancy in the rotational signature is less than $7 \%$ (respectively $\Delta \eta < 0.13$) and about $4-11 \%$ (respectively $\Delta \beta  \sim 0.04-0.11$) in the anisotropy. Results are reported in Table \ref{res}. 

With regard to a possible error in the BHB absolute magnitude (see \S ref{sec:absmag}), we re-analyse our sample accounting for shifts in $M_g$ of order of $\pm 0.20$ and find that both the rotational signature and the anisotropy are practically unaffected: we measure no systematic variations in either $\eta$ or $\beta$, with the maximum discrepancy being less than $4 \%$ for the rotation parameter and $5 \%$ for the anisotropy.  

We then contaminate our sample with $10 \%$ main-sequence A stars: we measure no appreciable shift in either the rotational signature or the anisotropy previously estimated. 

\begin{table}

$\begin{array}{|c||c|c||c|c|}\hline
& u-g>0.8 & & u-g>1.15 &\\
& \eta & \beta & \eta & \beta\\ \hline
\textrm{all} & 0.83 \pm 0.08  &-0.04 \pm 0.07 & 0.83 \pm 0.08 & -0.07 \pm 0.08 \\
\textrm{rich} & 0.75 \pm 0.12  & -0.24 \pm 0.11 & 0.79 \pm 0.12 & -0.35 \pm 0.11 \\
\textrm{poor} & 0.97 \pm 0.16  & 0.11\pm 0.11 &0.84 \pm 0.17 & 0.15 \pm 0.11 \\
\hline
\end{array}$\\

\caption{Estimates of the rotational signature and anisotropy of the Milky Way stellar halo with BHB stars from SDSS (DR9) selected according to two different cuts in the $(u-g)_0$ colour (the other filters stay as in \ref{FScuts}): the tighter cut in $(u-g)_0$ aims at excluding a suspected blue hook.}\label{res}

\end{table} 

Given the on-going debate on the value of $R_0$ and of the Sun's peculiar velocity \citep[see discussion in][]{sch12}, we test our results by considering the following variations in the radial position of the Sun: $R_0=8.0 \textrm{, } 8.2 \textrm{, } 8.5$ kpc and find no difference in either the rotation or anisotropy retrieved. The same holds for alterations in the U-component of the Sun's velocity that span values from 10 km/s up to 16 km/s. What does affect the rotational signature estimated through our analysis is the Sun's azimuthal velocity: raising it from the adopted $v_{c \odot}=220$ km/s to $v_{c \odot}=250$ km/s, makes the rotation signature $\eta$ rise by $0.5$ towards pro-rotation. However, the significance of this shift is more intuitive looking at the direct estimators: the $+30 \kms$ change in $v_{c \odot}$ translates into a shift in the rotation estimate of $< 10 \kms$ towards pro-rotation.

\section[]{Comparison with Deason et al. 2011}

D11 presented a study on the kinematics of the Milky Way stellar halo using SDSS BHB stars from DR7 and retrieved a metal-poor counter-rotating halo and a metal-rich pro-rotating halo. They identified BHB halo tracers by cutting in colour, surface gravity, effective temperature and position to remove the disk and the Sagittarius stream as we do in (\ref{FScuts}), except they also include stars with\footnote{Note that given the higher average surface gravity value in SDSS data releases prior to DR9 by $\sim 0.2$ dex \citep[][]{ahn12}, for DR7 the surface gravity cut in (\ref{FScuts}), reads: $2< \log g<3.5$ (see \S \ref{sec:daterr} for a discussion).}:

$$
\left\{ \begin{array}{l}
g \ge 18\\
3.5 \ge \log g <4.0\\
\end{array}\right.
$$

\subsection{Discussion of D11's selection criteria}

The inclusion by D11 of stars with $\log g \in [3.5,4.0]$ and $g \ge 18$ exposes them to the risk of contamination by non-BHB stars \citep[][]{sir04,lee08}. Therefore, the rotational signature retrieved with their sample is more prone to be biased: according to the literature value they buy their larger sample size with a contamination by misselected stars of $\sim 25 \%$ (compared to $10\%$ for our sample, see section 3) and the inclusion of faint stars makes their sample span distances from the GC up to 90 kpc, while we cover a sphere of radius $30-40$ kpc. Therefore the geometrical biases outlined in \S 3.1 will be more significant for the sample of D11 than for ours. 

We confirm that if we drop our tighter cuts in surface gravity and apparent magnitude, we retrieve the results of D11\footnote{With the sample drawn from SDSS DR7, filtered as prescribed by D11 and assuming $M_g=0.7$, we find the same two populations (a metal-poor one with $-3<\fe<-2$ and a metal-rich population identified by $-2<\fe<0$) and estimate $\beta$ and $\eta$ for them separately, according to the likelihood (\ref{MLH}). We confirm the trend presented by D11 of a pro-rotating metal-rich population versus a counter-rotating metal-poor one and match their numerical estimates of $\eta$ and $\beta$ with the exception for the anisotropy parameter for metal-rich stars. D11 find $(\eta,\beta)_{\rm rich}=(0.8  \pm 0.1,0.1 \pm 0.1)$ and $(\eta,\beta)_{\rm poor}=(1.4  \pm 0.1,0.2 \pm 0.1)$ while we find $(\eta,\beta)_{\rm rich}=(0.7  \pm 0.1,0.5 \pm 0.2)$ and $(\eta,\beta)_{\rm poor}=(1.5  \pm 0.1,0.3 \pm 0.1)$. We suspect this difference traces back to their undescribed handling of the singularity at $L=0$ by D11 and to the $15\%$ difference in sample size.}. Also, we try the heuristic exercise of perturbing the distances in the sample of D11 by $20 \%$ and verify that the inclusion of faint stars makes the rotational signature highly susceptible to distance errors. 

\begin{figure}

\epsfig{file=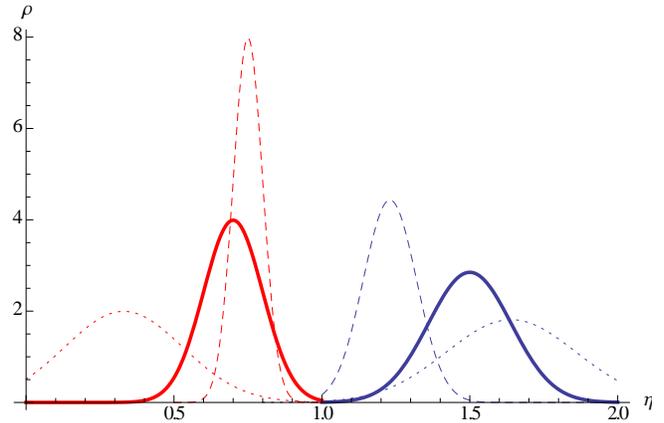,angle=0,width=85mm}

\caption{Probability density of $\eta$ marginalized over $\beta$ for BHB stars from DR7 (see 4.1): blue lines for metal-poor stars and red lines for metal-rich ones. Full lines represent D11's results, who assume that $M_g=0.7$ for each star in the sample. Dashed lines show how the rotational signatures change if the distances have been underestimated by $20 \%$, dotted lines represent the signatures associated with a distance overestimates of the same order. }

\label{dist_misest}

\end{figure}

We assume a misclassification of $\Delta M_g=\pm 0.40$, corresponding to a distance underestimate of $\pm 20 \%$\footnote{We remark that regardless of the theoretical nature of this exercise, these values for $\Delta M_g$ are not at all unrealistic (see Section 2).}. Fig. \ref{dist_misest} shows that for distance overestimates (dotted lines) the rotational signatures are noisier and the discrepancy between the metal-poor (blue lines) and the metal-rich (red lines) sample is exacerbated, while for distance underestimates (dashed lines) they shift towards non-rotation and the signal reduces.  On the other hand, when we challenge our sample with a perturbation in the adopted absolute magnitude of similar intensity, we measure no scatter greater than one sigma in the rotational signatures previously estimated. It seems helpful to remark that distance errors do not change the sign of rotation, but rather enhance (for distance overestimates) or shrink (for distance underestimates) the measured rotational signature: this is a pure geometrical consequence for measuring rotation via the component of velocity that at the limit of infinite distance is perpendicular to the direction of rotation. Therefore, if there is any rotational signal a distance bias will inflate it or reduce it, but if the system is non-rotating, errors in distance will have no effect on the estimated streaming motion. Distant stars are particularly sensitive to distance systematics for the effect of a bias scales with distance (see Fig. \ref{weight_scale}). This explains why our sample is so little affected by perturbations in the distance scale, while the signature found by D11 is not robust in these respects.

In conclusion, the rotation measured by D11 is driven by stars with $\log g \in [3.5,4.0]$ and $g \ge 18$. When we remove these stars, with the same analysis we measure no rotation. We note that this exclusion does not bias the properties of our sample: for example the distributions of $\vlos$ and of $T_{\rm eff}$ are unaffected by the tighter filters. The $g <18$ cut very reasonably shifts the peak of our distance distribution towards a shorter mean, but has no effects on the other properties. Further, this does not affect our conclusions since the signal claimed by D11 is constant with radius. The removal of $\log g \ge 3.5$-stars diminishes mainly objects from the metal-rich sub-sample and far less from the metal-poor sub-sample. Since the controversial signature resides on the metal-poor side of the sample, the signal should be enhanced by the altered proportions, but this does not happen.

Besides the concerns associated with the classification of these stars, if the signature measured with D11's more generous filter were a global property of the sample considered, removing \emph{random} elements from it would widen the spread around the mean, but keep the latter the same\footnote{Random is here understood as statistically non significant and the latter consideration is a consequence of the assumption that there are no spectral contamination in the population.}. On the contrary the comparison shows a shift in the means and a decrease in the formal errors, confirming the suspicion that the objects with $\log g \in [3.5,4.0]$ and $g \ge 18$ are non-BHB stars. 

\subsection{Consistency between model and data}

To assess whether there is any systematic in the analysis of D11 due to the more generous filters in apparent magnitude and surface gravity, we run the consistency check introduced  on their sample (see \S \ref{sec:geocheck}). In the subsequent subsection \ref{sec:522} we will trace back the detected inconsistency to a likely disc contamination.

\subsubsection{Spatial dependence of the rotational signal}

We bin $(l,b)$-space to have cells with useful statistics according to the $(l,b)$ distribution of BHB stars within SDSS: we attempt a regular grid and perform the analysis in each bin. We compute the formal error in each bin, and since they are all of the same order we consider the regular grid a fair choice. There is of course a geometrical effect one needs to account for: the sphericity of the reference frame causes no rotation at the poles ($b=\pm 90$) and maximum rotation on the galactic plane. However, given the significantly big bins chosen, this produces a smooth fall of the rotational signature measured from the plane up to the poles, rather than jumps. From one bin to the next the mean rotation velocity is therefore expected to be consistent within the error bars. The only place where we can have a real physical discontinuity in the rotational signature is the region which includes the GC.

\begin{figure}

\epsfig{file=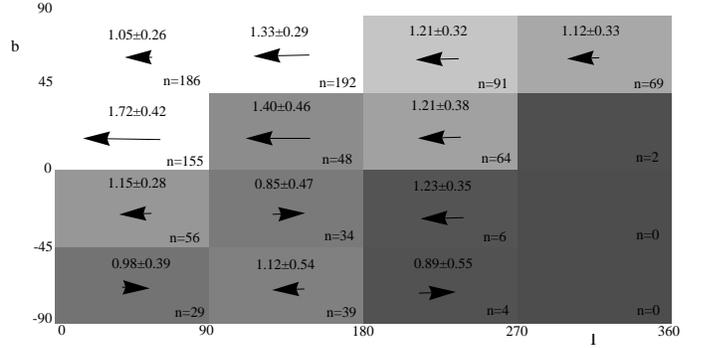,angle=0,width=90mm}

\caption{Analysis of the rotation parameter $\eta$ on metal-poor BHB stars from SDSS DR7 as selected in D11 and performed locally in $(l,b)$-space. Arrows pointing right indicate pro-rotation and vice versa for arrows pointing left. Their magnitude is a visual indication of the rotation strength while the actual value is reported above them together with the associated formal error. Bins are coloured according to the density of stars (reported in the bottom-right corner of every bin). The colour scale is normalized per each population. }

\label{DR7_lb_eta_map}

\end{figure}

\begin{figure}

\epsfig{file=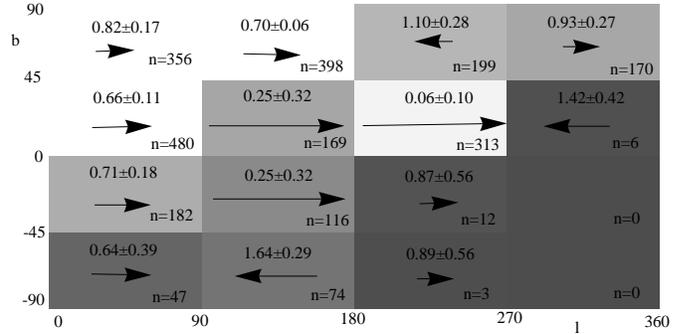,angle=0,width=90mm}

\caption{Same as in Fig. (\ref{DR7_lb_eta_map}) for metal-rich stars. Note the jump in signature between the bin centered in $(135,-68)$ and the one in $(135,-23)$. Or between the bin centered in $(225,23)$ and the one in $(225,68)$ bin. }

\label{DR7_lb_eta_map1}

\end{figure}

Fig. \ref{DR7_lb_eta_map} shows the bins in $(l,b)$-space with an arrow in each bin pointing in the direction of the recovered rotation for the metal-poor component. The arrow lengths are associated to the strength of the rotational signature, and the numbered values of $\eta$ are reported above the arrow. We see that the best values for the rotational parameter are consistent, within the errors, with a non rotating stellar halo ($\eta=1$). Fig. \ref{DR7_lb_eta_map1} shows the same analysis for the metal-rich population: here the situation becomes pathological as there are jumps from pro-rotating to counter-rotating motions. Compare for example the bin centered on $(135,-68)$ with the one on $(135,-23)$ or the bin centered on $(225,23)$ with the one on $(225,68)$. 

From this, we conclude that the model (\ref{model_df}) is unable to fit the sample of D11 and the parameters estimated using this model cannot be used to achieve any further understanding of the system. Indeed, if the model is a bad representation of the data, the parameters estimated from it, will be driven by the noise that arises from the discrepancy between the model and the true distribution, rather than by the real signal in the data. 

A second consistency check is presented in Appendix C, where we show that for a distribution function like (\ref{model_df}) and for fixed $\beta$, the best-fitting value for the entire population should be a linear combination of the best-fitting values of any sub-samples one breaks it into, where the coefficients are the fractional sizes of each sub-sample with respect to the total size. Our results satisfy this constraint, but the ones reported by D11 do not. From their Fig 4. $\eta_{\rm rich}=0.8$, $\eta_{\rm poor}=1.4$, $\eta_{\rm total}=1.1$. For fixed $\beta=0.1$, having $32\%$ metal poor stars and $60\%$ metal rich stars, one should have $\eta_{\rm total}=0.9$ (if one considers the 3549 stars) or 1.0 (if one previously  excludes from the total sample stars with ``no assigned metallicity'', i.e. $\sim 300$ stars).

The U-estimator (see 3.4.1) supports the conclusions of the two sanity checks just introduced: on the sample of D11 we measure an overall very weak counter-rotation actually consistent with no-rotation at all: $\langle v_\phi \rangle=4 \pm 10$ km/s. However, with this sample, the radial velocity of the Sun is offset by a bit more than $1$ sigma: $U_\odot=1.4 \pm 6.7$ km/s. 

\subsubsection{Likely disc contamination in D11's sample}\label{sec:522}

The prograde rotation measured by D11 on their metal-rich sample is linked to the spatial inconsistency of their data-model: Fig. \ref{DR7_lb_eta_map1} shows that the systematic prograde rotation of metal rich objects occurs at low latitudes and fades away at higher latitudes. A priori this would be consistent with either a disc contamination, the lower halo being set into stronger rotation by friction with the disc or the debris of an accretion event.

In an attempt to break the degeneracy, we vary the lower altitude cut. When we vary the z-cut from 4 to 6 kpc (cutting about $\sim 10\%$ of the sample) in the D11 sample, the $\vlos$-estimator confirms a statistically significant drop in the rotational signal of metal-rich stars from $34.1\pm7.4 \kms$ to $23.5\pm8.2 \kms$. At a cut of 14 kpc the $\vlos$-estimator has lost its significance by dropping to $12.5 \pm 16.6 \kms$. Hence we know that basically all the prograde signal is coming from stars close to the Galactic disc. How reliable is it?
To answer this, we look again at the 3D rotation estimators and compare to the other, cleaner datasets.

In the cleaner FS12 sample we find significantly slower rotation of $22.2 \pm 12.0 \kms $ for $|z| > 4 \kpc$ and a statistically insignificant signal of $11.6 \pm 12.6 \kms$ for $|z| > 6 \kpc$, while the X11 sample, which should have the lowest contamination with blue stragglers/main-sequence stars, has no prograde rotation at all (see Table \ref{res}). Hence, with increasing data quality the signal diminishes.

On the other side the 3D estimators do not display any rotation at all. In the most strongly rotating case of D11 we find:
\begin{eqnarray*}
\langle \pi^{-1} (\vlos) \rangle &=& (34 \pm 7) \kms  \textrm{,} \\
\textrm{but } \langle  v_\phi \rangle_{U-est} &=& (-2 \pm 10) \kms \textrm{,\quad }  \langle v_\phi \rangle  = (3 \pm 6) \kms
\end{eqnarray*}
 where the values are taken for $|z| > 4 \kpc$. This inconsistency could be the consequece of part of a stream in the sample, but even then its rotation wouldn't be necessarly prograde. Further, the 3D estimator $\langle v_\phi \rangle$ shows a hint of rotation in the near-disc region and interestingly gets into better agreement with $\pi^{-1} (\vlos)$ when we shorten all distances, while the $U$-estimator stays consistently at $0$. All this points to the simpler and therefore favourable hypothesis of a contamination with fainter disc objects.  

\noindent It is worth remarking that for the sample of D11, disc contamination is likely to exceed the geometric limits of the disc itself: the a priori comparison with the sample of \cite{x11} flagged half of the stars as probable non-BHB. If a blue straggler/main sequence star is mistaken for a BHB member, its distance is significantly overestimated, so disc contamination due to these objects can affect the BHB sample well above the actual disc limits.

In summary even on the D11 sample the 3D estimators refute any reliable evidence for a clean prograde rotation, while the difference between the estimators points to a contamination by disc objects. This does not mean that no prograde tendency in the near-disc region of the Galactic halo exists, but it is not detectable at the accuracy of the \cite{de11} sample or other samples.

\subsection{Rotation gradient}

\subsubsection{In metallicity}
The gradient in rotation that D11 measure between samples of different metallicity could at first sight be interpreted as support of the claim of a double-component stellar halo, even though D11 w.r.t. their result advocate the need of redefining the Sun's azimuthal velocity to $250 \kms$ and impute the rotation gradient to contamination from an accretion event.\footnote{We remark that drawing from the SDSS DR7 catalogue and applying the same cuts as in D11, lead to obtain a slightly different sample size with respect to D11. With the colour, gravity, temperature and height filter we have 5523 stars (D11, 5525) and when masking out Sgr stream and any close star ($r>10 \kpc$) the sample shrinks to 3500 stars (D11, 3549). Of these, 2525 fall in the metal-rich population (D11, 2125) and 975 in the metal-poor population (D11, 1135) even though we are using the same metallicity indicator \citep[][]{wil99}.}  
Besides the fact that with stringent quality cuts we do not measure any rotation discrepancy between metal-poor and metal-rich stars (see \S 4), we investigate if there is any further ground in the sample of D11 that could support the thesis of a double-component stellar halo. We have searched systematically for evidence of bimodality in other distributions and can report that no bimodality is evident in any of the following planes: $\fe$ versus any of $u-g$, $g-r$, $\log(g)$, $g$-mag or $v_\phi$; $v_\phi$ versus any of $g$-mag, $u-g$ or $r$; or $r$ versus $u-g$.

\subsubsection{In radius}

\begin{figure}

\epsfig{file=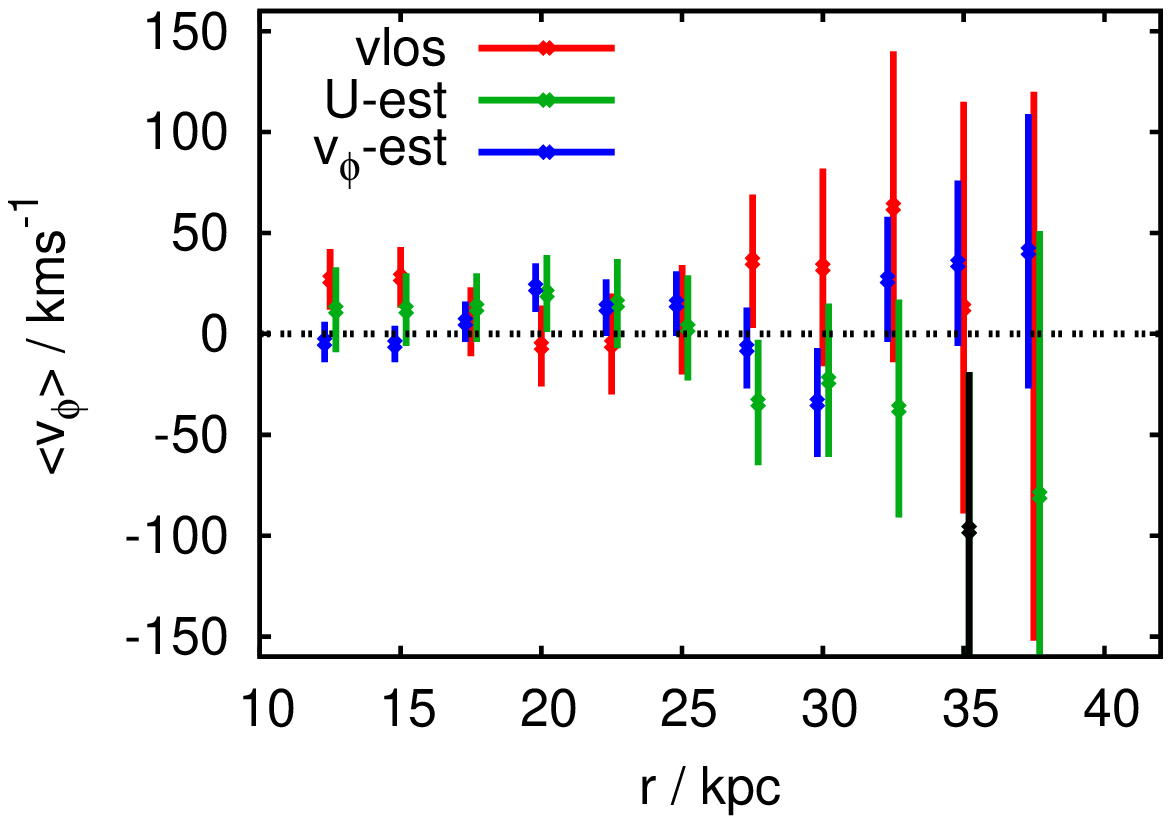,angle=0,width=85mm}
\epsfig{file=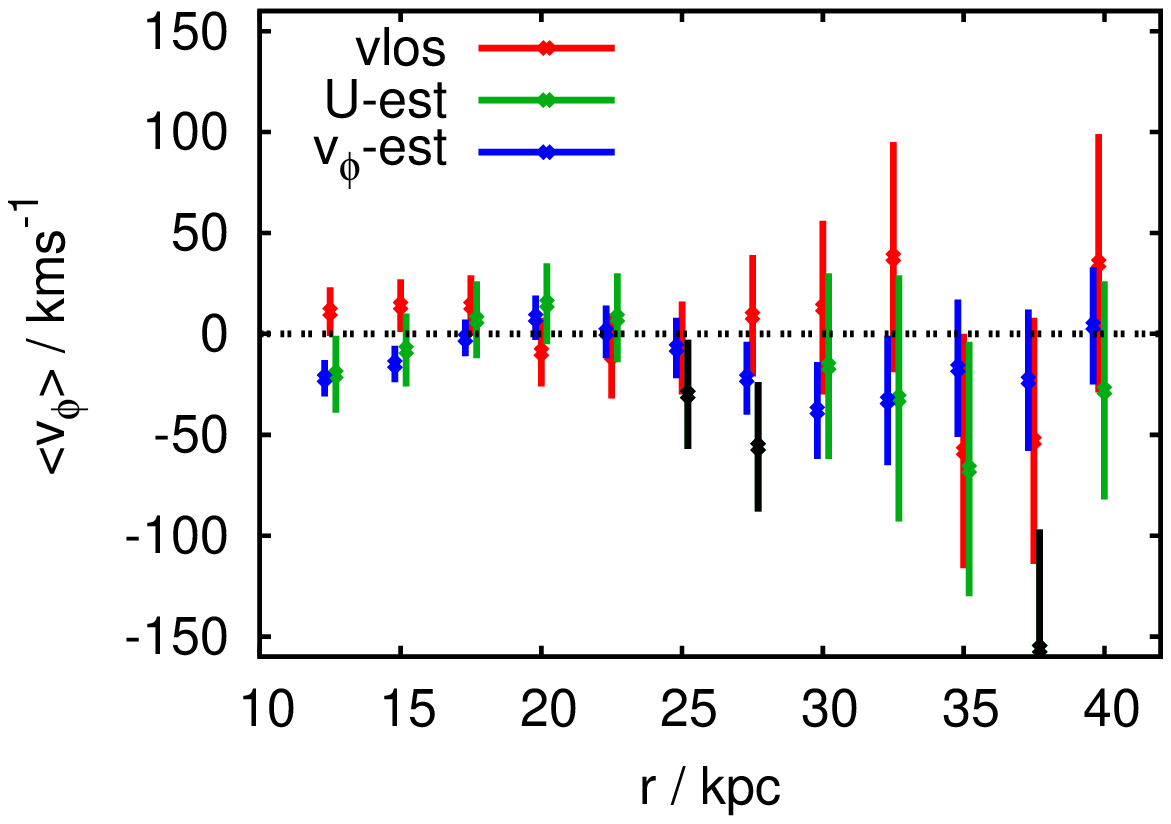,angle=0,width=85mm}

\caption{{\it Mean streaming motion as function of galactocentric radius for the FS12 sample (top panel) and stars in the X11 sample with $c_\gamma<0.95$ (bottom panel). In one bin out of eleven for FS12 and three out of twelve for X11, the $U$-est returned a solar value in disagreement with the literature: the level of systematics in those bins is therefore relevant and the rotation estimate cannot be trusted there (dark green points).}} 

\label{rotVSr}
\end{figure}

On the other hand, \cite{car07,car10} and \cite{kin07,kin12} suggest the existence of a double halo in terms of real space position: an inner halo dominant at 5-10 kpc and an outer (counter-rotating) halo dominant beyond 20 kpc. We investigated this possibility both with our samples and with the selection of D11, but find no evidence for such bimodality in either the radial density profile or in the radial distribution of the rotational signature (e.g. outer halo counter-rotating and inner halo pro-rotating). 
Fig. \ref{rotVSr} shows the mean-streaming motion as a function of galactocentric radius: we split the sample into $5 \kpc$-wide bins, values being shown every $2.5 \kpc$ so that every second value is independent. The top panel refers to the FS12 sample and the bottom one to the low $c_\gamma$ stars in X11 (see discussion in the following section): to determine the significance of the deviations from a zero constant rotation we run a \emph{p-test} on both trends. Assuming a certain null-hypothesis (model) is true, the p-value is the probability to measure a deviation from it as large as the one observed. The levels of significance to reject a null-hypothesis are most commonly taken to be $1-5 \%$. The p-value for the trends shown in Fig. \ref{rotVSr} confirms that the deviation from a  constant zero rotation is statistically insignificant for all estimators on all samples:\footnote{In our analysis we use 1 minus the cumulative distribution of the $\chi^2$-distribution to estimate the p-value. Note that because our bins overlap by half their size, we effectively have half the degrees of freedom we would have if the bins were all independent.}
$$
\begin{array}{c|ccc}
{\rm p-value} & \vlos{\rm-est} & U{\rm-est} & v_\phi{\rm-est} \\ \hline
{\rm FS12} & 0.53 & 0.75 & 0.50 \\ \hline
{\rm X11} & 0.80 & 0.82 & 0.26 \\ 
\end{array}
$$

\noindent In Table \ref{2pop} we report the actual estimates of the mean streaming motion for two spatial components, one of stars closer than $15 \kpc$ in galactocentric radius and one of stars further than that. All estimators show\footnote{The discrepancy between the numerical estimates reflect the different sensitivities of the estimators to the sample geometry. } a halo weakly rotating if at all and that there is no appreciable signal between the components either spatially or chemically selected. Note that the prograde rotation found by \cite{kaf} at small radii in their metal-rich sample of BHB stars drawn from the catalogue of \cite{x11}, is based on $\vlos$ information only: we find no grounds to support it looking at the more robust 3D estimators (see \S \ref{sec:daterr}).  Even though this could be the kinematic trace of a debris from an accreted galactic fragment, it is helpful to note that a plausible 10-15 $\kms$ aberration in $\vlos$ \citep[][]{sch12} would cancel any hint of rotation also in the $\vlos$ estimators, leaving basically unaltered the 3D estimators. Further, the discrepancy in the rotational signal that \cite{kaf} see between metal-rich and metal-poor stars,  is enhanced both by an inappropriate choice of distance scale, and by the kinematic bias afflicting high $c_\gamma$ stars in the sample of \cite{x11} (see  below).

\begin{table}
$\begin{array}{|c||c|c|c|}\hline
v_\phi (\kms) & v_\phi \textrm{-estim }  &\textrm{U-estim }   & \vlos\textrm{-estim } \\ \hline \hline
r < 15 \kpc & & & \\ \hline
& -3 \pm 8 & 7 \pm 19 & 22 \pm 13 \\
-2<\fe<0 & 10 \pm 11 & 9 \pm 24 & 31 \pm 16\\
-3<\fe\le -2 & -21 \pm 13  & 6 \pm 32 & 7 \pm 21 \\
\hline \hline
r \ge 15 \kpc & & & \\ \hline
& 7 \pm 7 & -4 \pm 11 & 4 \pm 11 \\ 
-2<\fe<0 & 12 \pm 10 & -6 \pm 15 & 14 \pm 15 \\
-3<\fe\le -2 & 0 \pm 11 & -1 \pm 18 & -8 \pm 17 \\
\hline
\end{array}$

\caption{We test the hypothesis of a double component stellar halo. With the BHB sample from SDSS DR9, the model-independent estimators of Section 2 measure no significant signal both between the kinematics of the closer stars as opposed to the furthest ones and when we discriminate for their metallicity.}

\label{2pop}
\end{table} 

\section[]{‪Critique of the signature in the Xue et al. sample‬}\label{sec:x11critique}

As in the case of \cite{de11}, the data-model pair for the X11 sample fails the consistency check introduced in \S 5.2, so we need to investigate the sample in detail. 

\begin{figure}

\epsfig{file=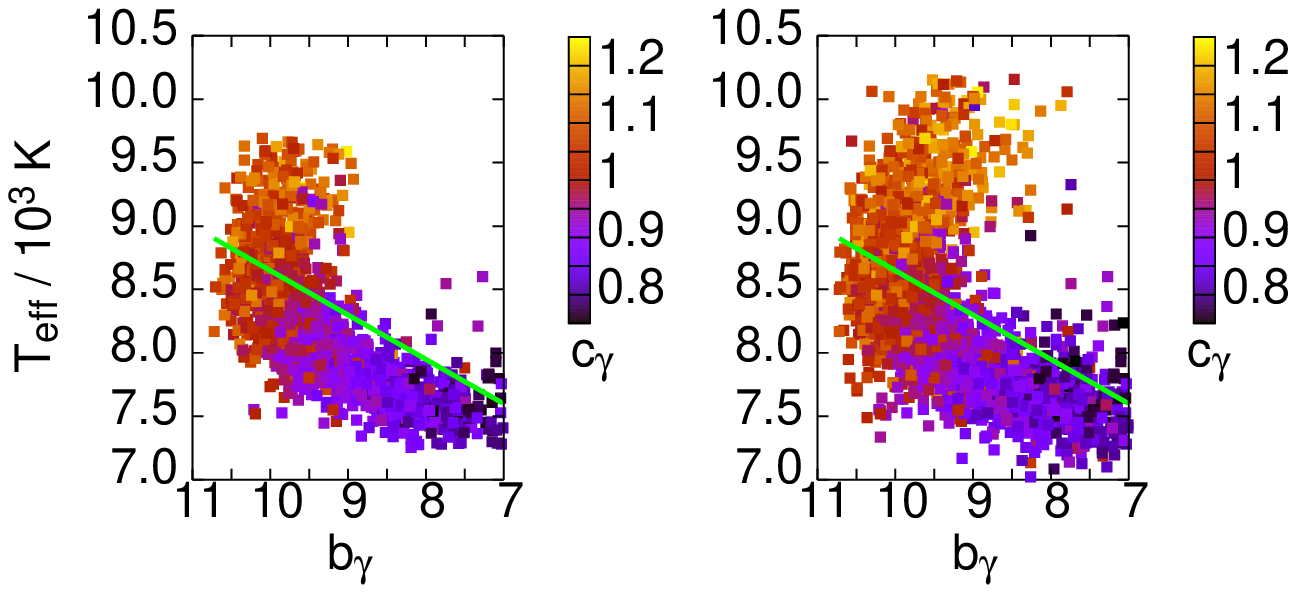,angle=0,width=85mm}
\epsfig{file=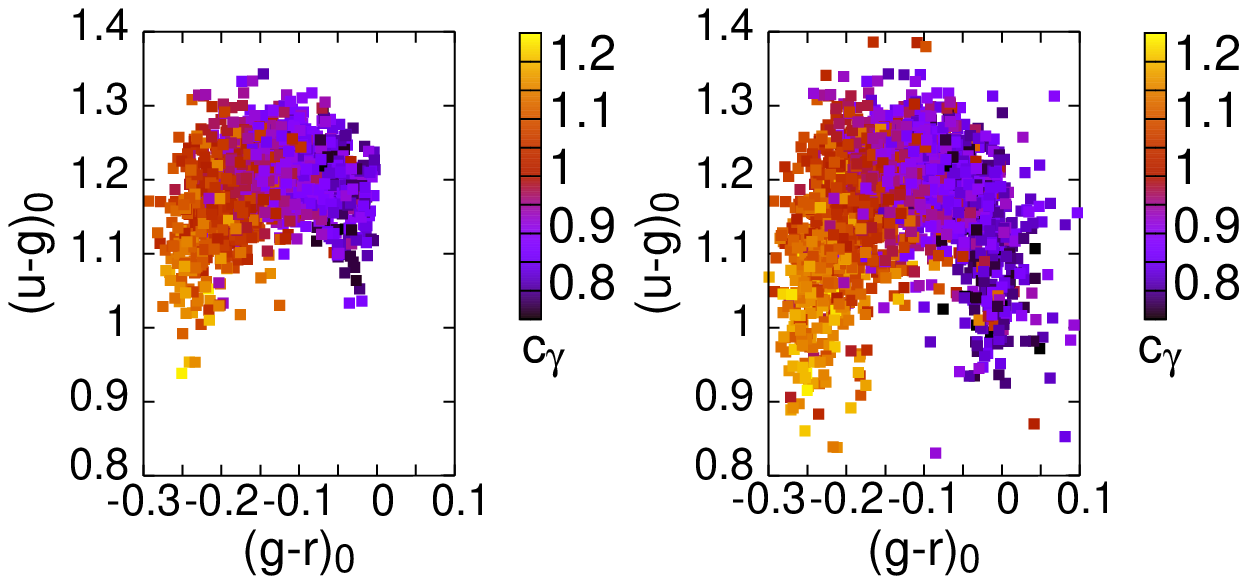,angle=0,width=85mm}

\caption{BHB stars that appear in both X11's sample and in the sample photometrically selected via (\ref{FScuts})  (left panels) and the full X11's sample (right panels). In the top panels we plot $b_\gamma$ versus $T_{\rm eff}$: the green line indicates what by eye appears to be a change in trend. X11's sample clearly enhances the region above the line: in the colour-colour plane (bottom panels) that is associable with the blue hook. The colour code reflects the value of the $c_\gamma$ Sersic parameter.}

\label{Teff_bgamma}

\end{figure}

In Fig. \ref{Teff_bgamma} we see that X11's sample strongly enhances the high temperature-high $b_\gamma$ region previously unexplored by the photometrically selected sample. Also, the relation between $T_{\rm eff}$ and $b_\gamma$ clearly changes in this region w.r.t. lower temperatures. Separating the two trends by eye (green line in Fig. \ref{Teff_bgamma}) the rotation discrepancy between metal-rich and metal-poor stars below the boundary disappears (metal-rich : $\eta=0.76 \pm0.15$;  metal-poor: $\eta=0.94 \pm 0.15$). On the other hand, the stars above the line show a strong counter-rotation as well as a signal of $\Delta \eta \sim 0.66 \pm 0.19$ (metal-rich: $\eta=1.11 \pm0.06$; metal-poor: $\eta=1.77 \pm 0.18$).

The few stars above the line in the photometrically selected sample show the same trend. Their strong under-representation ($20 \%$ versus $50 \%$ in X11) explains the discrepancy towards X11 (left panel in Fig. \ref{Teff_bgamma}).

\subsection{Kinematics of high $c_\gamma$ stars}

\begin{figure}

\epsfig{file=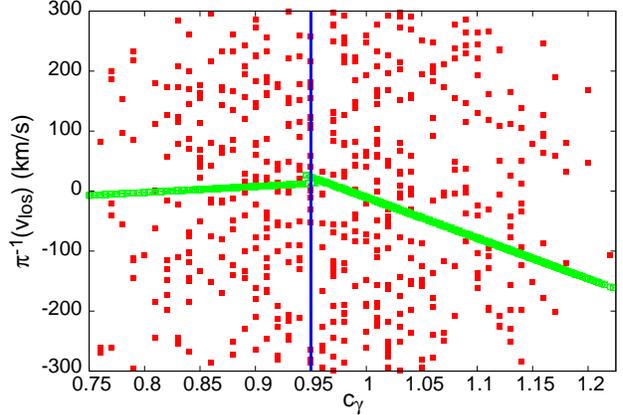,angle=0,width=85mm}

\caption{Deprojected l.o.s. velocity for metal-poor BHB stars spectroscopically selected by X11 as function of the line indicator $c_\gamma$. Only stars with $\vdep \in [-300,300] \kms$ are shown (red squares), but the least square fit to the $\vdep$-trends plotted here with green lines have been obtained on the full sample.} 

\label{cgammavdeproj}

\end{figure}

This dichotomy can be linked to the steepness parameter $c_\gamma$, which is a line indicator characteristic of luminosity type. While it does not directly select on kinematics we see a strong systematic difference within the metal-poor stars between high and low $c_\gamma$. 
In Fig. \ref{cgammavdeproj} we plot the de-projected l.o.s. velocities as a function of the line indicator $c_\gamma$ (red squares) and show the least square fit to $\vdep(c_\gamma)$ in the two regions (green lines). We do not apply any kinematic cut in fitting the $\vdep$-trends, but for the plot we limit $\vdep \in [-300,300] \kms$ to make the fit difference in $c_\gamma$ better visible.
Metal-poor stars with $c_\gamma \ge 0.95$ display a significant retrograde signal that is absent in the low $c_\gamma$ stars. We note that these stars are not more distant than stars with $c_\gamma <0.95$; indeed the density distributions of the two sub-samples show no appreciable differences. The metal-rich stars do not seem to suffer from  this systematic.

\begin{table}
$\begin{array}{|c||c|c|c|c|}\hline
v_\phi (\kms) & \eta & v_\phi \textrm{-est}  &\textrm{U-est}   & \vlos\textrm{-est} \\ \hline \hline
c_\gamma<0.95 & & & & \\ \hline
{\rm all}& 0.90 \pm 0.17 & -13 \pm 8 & -20 \pm 16 & 10 \pm 11 \\
{\rm rich} & 0.93 \pm 0.16&  -13 \pm 11 & -27 \pm 21 & 11 \pm 14\\
{\rm poor} & 0.87 \pm 0.19& -12 \pm 13  & -11 \pm 24 & 8 \pm 17 \\
\hline \hline
 c_\gamma \ge 0.95 & & & & \\ \hline
{\rm all}& 1.11 \pm 0.13 & -2 \pm 7  & -7 \pm 13 & -10 \pm 9 \\ 
{\rm rich} & 0.91 \pm 0.10 & -5 \pm 8 & -10 \pm 16 & 8 \pm 11 \\
{\rm poor} & 1.4 \pm 0.19 & 2 \pm 10 & -2 \pm 21 & -36 \pm 14 \\
\hline
\end{array}$

\caption{Estimates of the rotational signature of BHB stars spectroscopically selected by X11. We find that stars with high $c_\gamma$ present a pathological correlation between line cuts and kinematics and reproduce the popular result of a metal-rich pro-rotating halo and a metal-poor counter-rotating halo. Unbiased stars instead confirm our findings of a weakly pro-rotating or non-rotating at all halo and the absence of signal between populations with different metallicity.}

\label{cgamma}
\end{table}

\begin{figure}

\epsfig{file=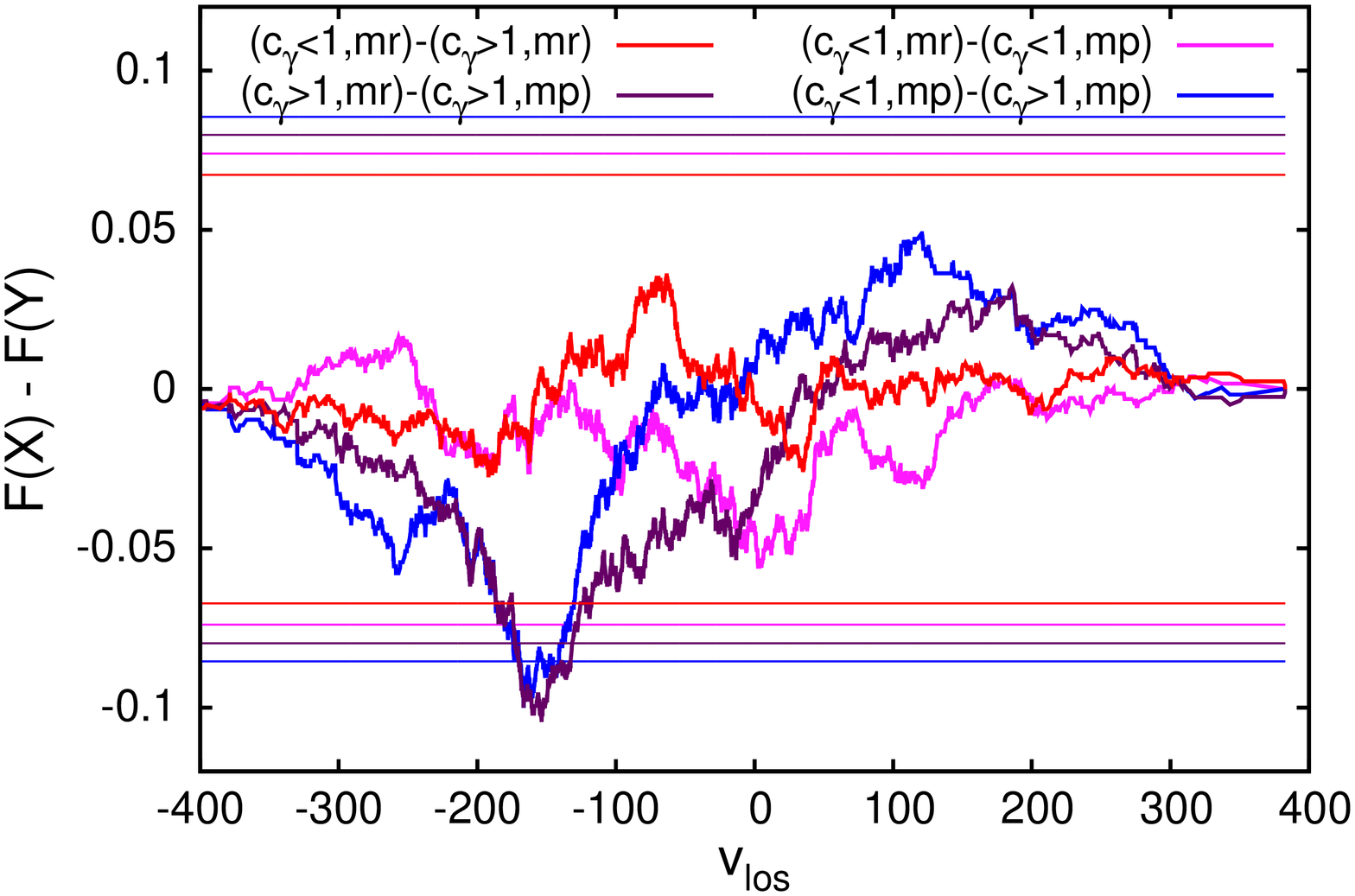,angle=-90,width=85mm}
\epsfig{file=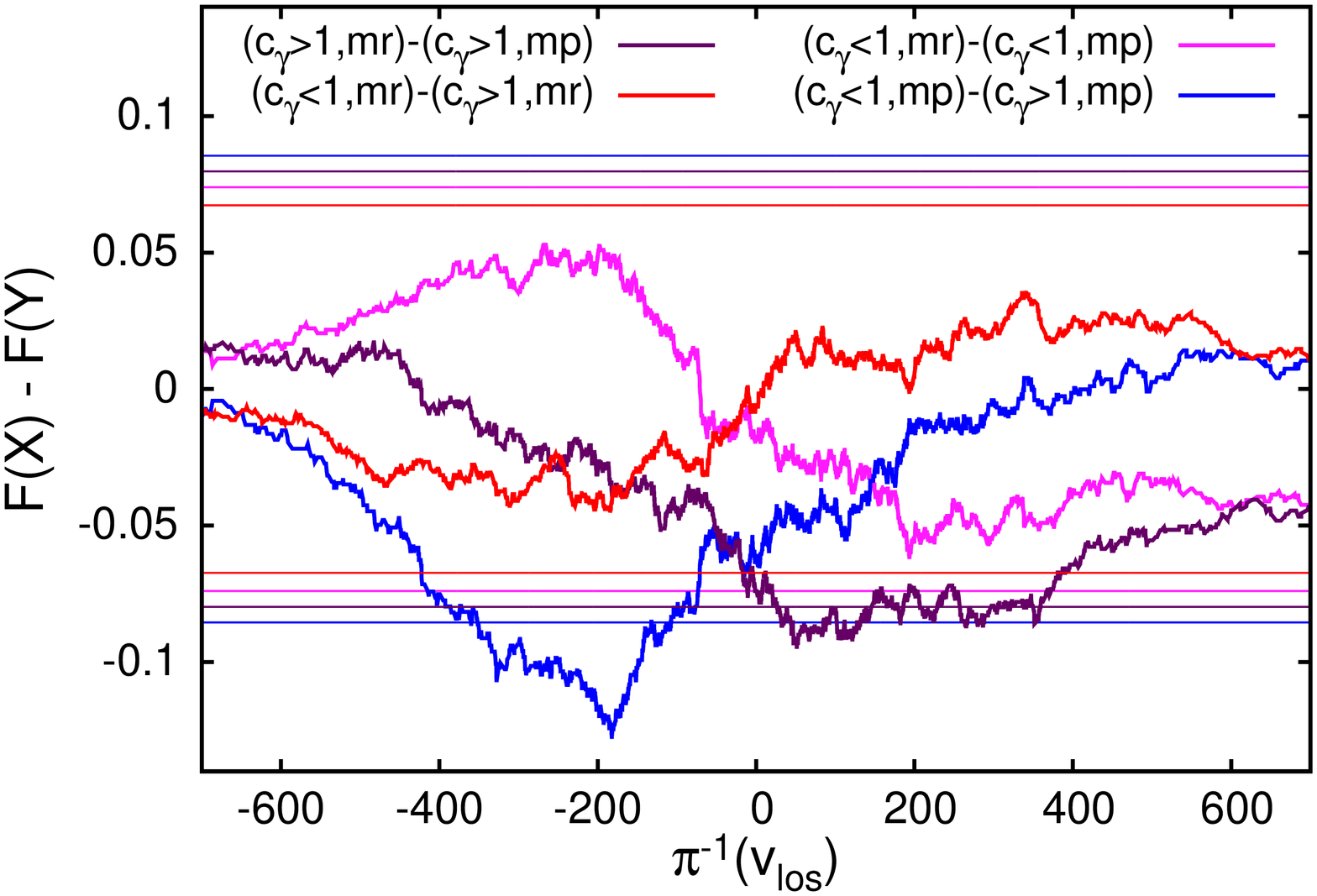,angle=-90,width=85mm}

\caption{Differences in the cumulative distributions of l.o.s. velocity (upper plot) and de-projected $\vlos$ (bottom plot) between sub-samples of different metallicity in two $c_\gamma$ bins. Red lines represent the discrepancy  between metal-rich stars with different $c_\gamma$, blue lines the same but for metal-poor stars. The magenta lines depict the difference between the velocity distributions of metal-rich and metal-poor stars with high $c_\gamma$, the purple lines the same but for stars with low $c_\gamma$. The horizontal lines are associated with the corresponding $5 \%$ significance levels of the Kolmogorov-Smirnov test.}

\label{KS_plot}
\end{figure}

We bisect our sample in both in $c_\gamma$ and $\fe$ and for each of the four sub-samples determine the cumulative distributions in $\vlos$ and $\vdep$. We then determine the Kolmogorov-Smirnov statistics to ask whether the former distributions can be drawn from a common underlying population. Fig. \ref{KS_plot} shows Kolmogorov-Smirnov tests on the discrepancy between the distributions of $\vlos$ (upper panel) and $\vdep$ (lower panel) for the four sub-samples. We plot the difference in the cumulative probability distributions of metal-poor and metal-rich stars in the two $c_\gamma$ bins (low $c_\gamma$: magenta line; high $c_\gamma$: purple line) as well as the difference between sub-samples of same metallicity, but different $c_\gamma$ (metal-rich stars: red line; metal-poor stars: blue line). For each test we show the $5 \%$ significance threshold with horizontal lines of the same colour. The discrepancy between two distributions is statistically significant if this threshold is crossed.

Indeed, we find that the most significant deviation occurs between metal-poor and metal-rich stars with $c_\gamma \ge 1.0$ (or, almost equivalently, for $b_\gamma >9.5$) and between metal-poor stars with different $c_\gamma$. By contrast, stars with $c_\gamma<1.0$ present no kinematic signal between sub-samples of different metallicity. In addition the significance of the signal between metal-poor stars in different $c_\gamma$ bins has exactly the same structure as the one between metal-rich and metal-poor stars with high $c_\gamma$ (and same significance).
Therefore, the Kolmogorov-Smirnov test confirms that the signature in the X11 sample is indeed caused by high $c_\gamma$ metal-poor stars. This inconsistency within the metal-poor sub-sample, raises severe doubts about the reality of the observed differences.\footnote{When we opt for the pipeline adopted metallicity instead of the WBG metallicity, the discrepancy remains and in fact is exacerbated.}

Even though with the current data we cannot rule out the possibility that these stars are associated with a real sub-structure in the halo, the $\vlos$-estimator (see Section 2.2.1) measures a non-zero gradient on these stars, pointing again to a pipeline problem.

These stars ($\sim 500$) are hot and very metal-poor so their spectra harbour nearly no metal lines that can support a radial velocity measurement, so this anomaly occurs in exactly the group of stars where we have the least trust in the velocity measurement. As can be seen from Table \ref{cgamma}, neither of the two 3D estimators show an appreciable retrograde signal. 
Proper motions seem not to contribute to the rotational signal and the weak one measured appears to come from $\vlos$ alone. In addition the $V$-$W$ estimator measures an underestimate in distance of $f=-0.10 \pm 0.09$ for these stars. Since there cannot be a significant contamination with more luminous stars (a contamination with blue stragglers or blue hook stars would give the opposite sign), this points to unusual kinematics in the vertical direction rather than a simple retrograde motion.

In conclusion, we find that metal-poor stars with high $c_\gamma$ do show a retrograde signal and we identify them as the source of the line-of-sight velocity distribution anomaly observed by \cite{bee12,hat,kaf}. By their inconsistency with their low $c_\gamma$ counterparts we flag this either as a pipeline problem or a special structure rather than as a second smooth component.\footnote{We also note that the discrepancy in metallicity distribution that \cite{bee12} advocate as evidence for a double-halo arises when comparing stars closer and further than $10 \kpc$.  The shift of $0.3$ dex is purely due to stars between 5 and 10 kpc, which still have a non negligible probability of being disk contamination (hence why both in \cite{de11} and here they are tossed out).
The tangentially biased metal-poor halo found by \cite{hat} on the other hand, is a consequence of the pseudo-rotation of high $c_\gamma$ stars, but the robustness of their measurement is also undermined by neglected treatment of possible contaminations. A minuscule fluctuation arising from a distance misestimate for example, can be enhanced to arbitrary level profoundly biasing the inferred anisotropy, hence the duty to quantify their vulnerability to distance errors and to deviations from the assumption of sphericity.}
If it were real, this structure would predominantly contain only hot BHB stars and have large systematic $W$ velocities across the sky while not showing a consistent retrograde motion. On the other hand, trustable stars (low $c_\gamma$) confirm our previous conclusions of a weakly pro-rotating or non-rotating halo irrespective of metallicity. Equivalent behaviours emerge when splitting the sample according to a line cut associated with $b_\gamma$.

\section{Conclusions}

We have analysed the kinematics of BHB stars drawn from SDSS/SEGUE (DR7, DR8, DR9), for which the only reliable measured component of velocity is $\vlos$, using both a selection via photometry and the sample of \cite{x11}. Using a very simple model and three model-independent estimators we find no evidence for rotation. Also, we measure no discrepancy in rotation between metal-poor and metal-rich stars. Following \cite{car07,car10} we look for a discrepancy in the kinematics of the inner and an outer halo, but find none in any of the possible selections. In both samples we trace back previously claimed rotational differences between different metallicity bins to stars that don't pass stricter quality criteria.

Our analysis is not restricted to the simple double power-law model used by \cite{de11}, but relies on three direct model-independent estimators: one uses the de-projected $\vlos$ and two make use of the full 3D motion. The latter require knowledge of proper-motion and while individual BHB stars are too remote to have sufficiently accurate proper-motion estimates, the errors being comparable to the signal, on the entire sample proper motions prove useful by large-number statistics. We checked that the 3D-estimators are robust both against change in the used Sloan data release and against systematics of the magnitude of the correction derived in \cite{sch12}. Systematics of higher magnitude would bias the 3D-estimators differently: the fact that they agree and that we recover the correct radial reflex motion $U_0$ confirms that the systematic floor is within the formal errors and the proper-motions useful. Moreover, the use of the 3D-estimators is crucial to test the results from the 1D-estimates that are highly vulnerable to pipeline systematics. A systematic shift in line-of-sight velocities causes a larger shift in the measured rotation, since the errors are amplified by the geometric factors and only partly balanced by having both sides of the sky. Further, a non-zero ``rotation" in the one-dimensional estimators does not prove a true retrograde or prograde net motion, but can be caused by failure of the assumption of zero mean $U$ and $W$ component.

Model-wise, overestimates in distance cause velocity space to shrink so certain stars are pushed out beyond the escape velocity with the consequence that the correct model assigns them zero probability. In this case the stars are eliminated from the group that contributes to estimates of the population's kinematics. This effect is not mirrored for distance underestimates, making the bias more severe. Unless observational uncertainties are appropriately folded with the model, again the 3D estimators are a necessary confirmation of any claim of net rotation.

We have proposed two general and one model-specific a posteriori consistency tests. The main idea being that a true rotational
signature in the sample should be seen consistently in any chosen portion of the sky. 
Re-analysing the photometrically selected sample of \cite{de11} we can recover their difference between prograde metal-rich and retrograde metal-poor populations. However, the signal is inconsistent across the sky and among the different estimators. In particular, the weak prograde signal in the metal-rich subsample shrinks considerably when we increase the geometric disk cut from $|z|>4$ to $|z|>6$ kpc. From the point of theory, it is plausble that due to accretion or frictional effects with the disc a confined region in the Galactic halo could display a mildly prograde rotation. However, the observations point to a disc contamination, given that the signal strongly weakens for the cleaner samples and that the 3D estimators do not confirm the rotation. We note that 
the additional constraints only reduce the level of contamination, while they do not a priori bias the inferred kinematics; when we repeat the same analysis on X11 we measure in fact homologous trends (see below). 
We further argue that the anisotropy measured via the adopted model is unreliable due to the unphysical $L^{-2 \beta}$ term in the distribution function, which for $\beta<0$ produces a bimodal distribution in the azimuthal velocity component.

We repeat our analysis on the spectroscopically selected sample of \cite{x11}. On the intersection of the two samples we do not find any hint for rotation. However, on the entire X11 sample we detect a weak offset in $v_\phi$ between metal-poor and metal-rich stars. While there is no detectable difference between the velocity distributions of metal-poor and metal-rich stars at low $c_\gamma$, we identify the metal-poor stars with high $c_\gamma$ as the source of this discrepancy. A Kolmogorov-Smirnov test reveals that the $\vlos$ distribution of these stars is not only offset significantly from that of metal-rich stars, but also from that of their metal-poor counterparts with low $c_\gamma$. Furthermore the retrograde behaviour in $\vlos$ is not confirmed by the 3D estimators, and anomalous $W$ velocities are detected. In fact these systematics are confined to metal-poor hot BHB stars that have very weak to non-detectable metal lines, which are crucial for reliable $\vlos$ determination. 

There are two possibilities: either this is a pipeline problem or these stars belong to a very metal-poor population containing only hot BHB stars. A systematic shift in line-of-sight velocities is not unlikely, given that the determination of line-of-sight velocities is very difficult in the smooth and nearly metal-line free spectra of hot, metal-poor BHB stars and that the average motion of stars towards the Galactic North and South pole indicates a significant pipeline systematic that increases towards blue, metal-poor stars in general.
If the component were real, it would not be in smooth retrograde motion, but rather likely form a dispersed stream-like structure with large systematic $W$ velocities.  High resolution spectral analysis of the suspected culprits will be required to decide if the SEGUE $\vlos$ are trustable for those stars, and if so, to identify the peculiar abundance pattern that can be expected for such a unique metal-poor structure. 

We tested our results against a variety of sources of uncertainty such as selection criteria, absolute magnitude estimate, contamination by main-sequence A stars, the Sun's parameters, the effect of reddening from the disk\footnote{We studied the effect that the choice to cut for $|z|<4$ kpc has on our results. We consider different cuts at $|z|=1,2,3,5,10$ kpc and observe no change in the signatures retrieved with our sample.} and distance errors up to $20 \%$. The results hold and prove to be susceptible only to variations in the Sun's circular velocity, for which we provide none the less a trend: changing $v_{c \odot}=220$ km/s to $v_{c \odot} =250$ km/s shifts the rotational signature towards pro-rotation by less than 10 $\kms$. 

The failure of the model used in this work teaches us that significant efforts have to be devoted to the creation and selection of appropriate models and strategies for fitting them to data. So far in fact, there is no a priori way to guess which characteristics a model should possess to fit satisfactorily the data and in fact, there is no agreed general definition of goodness of fit. There are ways to compare a model's ability to represent a certain sample (e.g. by computing the likelihood of the data given the model), but even with the best model among the ones available, there is no way to assess its intrinsic uncertainty.

We believe this area is still largely unexplored and should be considered the key step to cover for any meaningful analysis to be pursued.

\section*{Acknowledgments}
We thank James Binney and Heather Morrison for useful discussions and helpful comments on a draft of this paper. We further thank Alis J. Deason and Wyn N. Evans for providing the details of their analysis in D11. F.F. acknowledges financial support from the Science and Technology Facility Council (UK) and from Merton College, Oxford. R.S. acknowledges financial support by NASA through Hubble Fellowship grant $HF-51291.01$ awarded by the Space Telescope Science Institute, which is operated by the Association of Universities for Research in Astronomy, Inc., for NASA, under contract NAS 5-26555. Funding for SDSS-III has been provided by the Alfred P. Sloan Foundation, the Participating Institutions, the National Science Foundation, and the U.S. Department of Energy Office of Science. The SDSS-III web site is http://www.sdss3.org/

\appendix

\section*{Appendix A: MCMC chain convergence and test of implementation}
\subsection{Sampling parameter space}

We now explain how we retrieve ($\eta,\beta$) out of the log-likelihood (\ref{MLH}).

We construct a MCMC chain in parameter space, where the probability density is (\ref{MLH}). When can such a chain can be considered to have converged to a realisation of the model probability density? We establish convergence using the criteria by \cite{ver07}: given $T$ chains of length $2 N$, each with a different starting point, one estimates the chain-to-chain variance $B$ and the average variance of each chain $W$, which is an underestimate of the true variance:

\begin{eqnarray}
B&=&\frac{1}{T-1}\sum_{j=1}^{T} (\bar{\theta}^j-\bar{\theta})^2 \\
W&=&\frac{1}{T(N-1)} \sum_{j=1}^{T} \sum_{i=N+1}^{2 N}  (\theta_i^j-\bar{\theta}^j)^2,
\end{eqnarray}

\noindent where $\theta_i^j$ is the point in parameter space in position $i$ of chain $j$, $\bar{\theta}^j$ is the mean of the chain $j$, $\bar{\theta}$ is the mean of all the chains and the burn-in time $N$ is half the length of the chains.

An overestimate of the variance is
\begin{equation}
V=\sigma^2+\frac{B}{N T}=\frac{1}{N} \left[(N-1) W+B\left(1+\frac{1}{T}\right) \right],
\end{equation}

\noindent where $\sigma^2$ is the true variance. \cite{hea10} suggests that for convergence one should require $\frac{V}{W} \sim 1.01-1.1$: the precision reported is empirical and does not follow from a rigorous proof. We will therefore opt for a much more conservative threshold. 

Here, to avoid numerical issues when dividing quantities that are potentially null,  we adopt the following equivalent criterion:
\begin{equation}
\frac{V}{W}=(1+\varepsilon) \Rightarrow V-W=\varepsilon W,
\end{equation}

\noindent where $\varepsilon>0$ and
\begin{equation}
V-W=B(1+\frac{1}{T})-\frac{W}{N}.
\end{equation}

$T$ chains are therefore run separately but at the same pace: every time a new step is added to the chains, $W$ and $B$ are re-computed, and if the convergence criterion above is not yet satisfied, the chains keep growing. This avoids the need for an a priori fixed chain length and burn-in time. In our code we used: $T=2$, $\varepsilon = 10^{-3}$ and achieved convergence for $N$ typically of order $\sim 10^2-10^3$. 

\subsection{Test of implementation}

We test the implementation of likelihood analysis just introduced against $12$ mock samples: we generate a sample by MCMC sampling phase-space ($4 \times 10^4$ points) according to the distribution function in equation (\ref{model_df}) and random parameters $(\eta,\beta) \in [0,2]\times [-0.5,0.5]$. The reliability of the sample obtained is checked by comparing the density profile and rotational signature derived from star-counts of the synthetic catalogue produced with the theoretical profiles, revealing perfect agreement. 

\begin{figure}

\epsfig{file=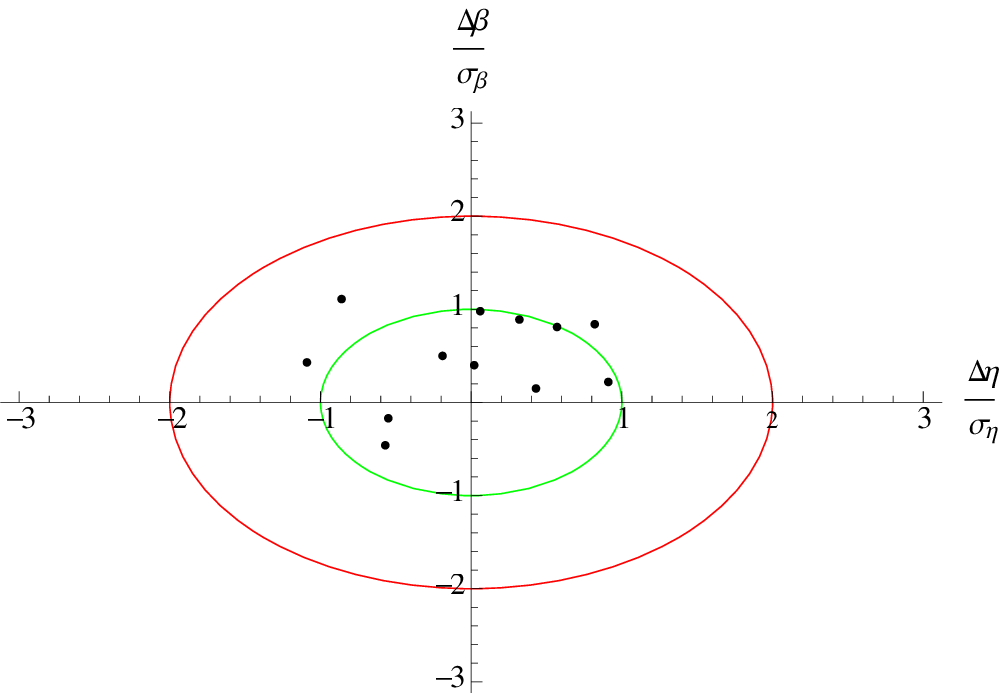,angle=0,width=80mm}

\caption{Performance of the MCMC fitting technique: the plot shows the discrepancy between the actual parameters and the fitted ones, weighted by the formal error. The fit is performed by MCMC sampling parameter space using the likelihood of the pseudo-data given the model. The green circumference represents the one $\sigma$ contour, the red one the two $\sigma$ contour.}

\label{syn_cat_proof}

\end{figure}

Subsequently, we determine the probability density of the model by MCMC sampling $(\eta,\beta)$-space using the likelihood of the pseudo-data given the model. The original parameters are recovered within one sigma as demonstrated in Fig. \ref{syn_cat_proof}. From the distribution of the deviations we also infer that the error estimates are reasonable.

\section*{Appendix B: Sample Selection}

The following SQL query was used to drawn data from the SDSS archive:

\noindent\begin{minipage}[t]{0.90\linewidth}SELECT\\
sp.PLATE,sp.FIBERID,sp.specObjID,ph.rerun,\\
,sp.TARGETSTRING,sp.mjd,sp.flag,sp.fehadop,sp.loggadop,\\
sp.fehadopunc,sp.FEHWBG,sp.TEFFWBG,sp.LOGGWBG,\\
ph.psfmag$_{\rm u}$,ph.extinction$_{\rm u}$,ph.psfmag$_{\rm g}$,ph.extinction$_{\rm g}$,\\
ph.psfmag$_{\rm r}$,ph.extinction$_{\rm r}$,ph.psfmag$_{\rm i}$,ph.extinction$_{\rm i}$,\\
ph.psfmag$_{\rm z}$,ph.extinction$_{\rm z}$,ph.l,ph.b,ph.ra,ph.dec,\\
m.pmL,m.pmB,m.pmRa,m.pmDec,m.pmRaErr,m.pmDecErr,\\
m.sigRa,m.sigDec,m.nFit,m.dist22,m.match,\\
sp.ELODIERVFINAL,sp.ELODIERVFINALERR,sp.SNR,\\
sp.TEFFADOP,sp.TEFFADOPUNC\\
FROM sppParams sp\\
JOIN PhotoObjAll ph on sp.bestobjid = ph.objid\\
JOIN ProperMotions m on m.objid = ph.objid\\
WHERE sp.sciencePrimary = 1\\
AND ph.CLEAN=1\\
AND sp.TARGETSTRING = 'BHB'\\
\end{minipage}

The masking of Sgr is performed in right ascension-declination space ($\alpha, \delta$) according to the following polygon kindly provided by the authors of D11 on request:

\begin{eqnarray*}
\alpha_1 &=& [245, 225, 211.5, 210, 208, 200.5, 199.3, 197.5, 197,\\
& & 180, 160, 140, 110, 113, 140, 160, 180, 202, 225, 245],\\
\delta_1 &=& [-3, 8, 14.2, 12.5,
16, 19.5, 16.5, 16.9, 21.5, 28, 33,\\
& & 35, 38, 15, 13.5, 11, 6, -3, -3, -3]; \\
\cup \textrm{ } \alpha_2 &=& [10, 50, 70, 30, 10] , \textrm{ } \delta_2 = [-10, 20, 10, -20, -10].\\
\end{eqnarray*}

\section*{Appendix C}

{\bf Theorem:}
 Let the sample of stars be $X=\{x_{(1)},\ldots,x_{(N)}\}$ divided in the sub-samples $X=X_1 \cup X^1=\{x_1,\ldots,x_m\}\cup\{ x^1,\ldots,x^n\}$. If $P(\eta|X)$ is the probability of the model relative to the parameter $\eta$ given the data and $\eta_{best,X}$ is the value that maximize it for the type of distribution function under consideration:

$$
\eta_{best,X}=\frac{card(X_1)}{card(X)} \eta_{best,X_1} +\frac{card(X^1)}{card(X)} \eta_{best,X^1} .
$$

{\bf Proof:}

Since $f \propto 1+ (1-\eta) \tanh(\frac{L_z}{\Delta}) $, assuming $\Delta \ll L_z$ and at fixed $\beta$ (such that the normalization constant, independent of $\eta$, is $n_c$):

$$
P(\eta|X)=\prod_{i=1}^{N} P(\eta|x_{(i)})= \frac{\prod [1+(1-\eta)\tilde{H}(x_{(i)})]}{n_c^N}
$$

where $\tilde{H}(x)=2 H(x)-1$ and $H(x)$ is the Heaviside function. Then one can split the above in the positive and negative contributions, obtaining:

$$
P(\eta|X)=\prod_{i=1}^{N^-}\frac{\eta}{n_c} \cdot \prod_{i=1}^{N^+}\frac{2-\eta}{n_c}
$$

where $N^+$ is the number of stars with $L_z>0$ and the opposite for $N^-$. The probability is maximum for $\frac{\textrm{d} P(\eta|X)}{\textrm{d}\eta}=0$, which happen for $1-\eta=\frac{N^+ - N^-}{N}$. The same holds for the two sub-samples of $X$: calling $m^+$ the number of stars with positive $L_z$ that belong to $X_1$ and $n^+$ the ones belonging to $X^1$ and similarly for stars with negative $L_z$, we have:

$$
\eta_{best,X_1}=\frac{2 m^-}{m} \textrm{  \qquad  and  \qquad } \eta_{best,X^1}=\frac{2 n^-}{n}
$$

with $N^+=m^+ + n^+$ and $ N^-=m^- + n^-$. We conclude, showing that $\eta_{best,X}$ can be expressed as follows:

$$
\eta_{best,X}=\frac{2(m^- + n^-)}{N}=\frac{2 m^-}{m} \cdot \frac{m}{N}+\frac{2 n^-}{n}\cdot \frac{n}{N}
$$

This completes the proof. 
\label{lastpage}
\end{document}